\documentclass{aa}
\usepackage{natbib}
\usepackage{multicol}
\usepackage{siunitx}
\bibpunct{(}{)}{;}{a}{}{,}
\usepackage{graphicx}
\graphicspath{{./figures/}}
\usepackage[varg]{txfonts}
\usepackage{pdfpages}
\usepackage{booktabs}
\usepackage{tabularx}
\usepackage{xspace}
\usepackage[normalem]{ulem}

\newcommand{\gcc}{\ensuremath{\mathrm{g} \, \mathrm{cm}^{-3}}}

\newcommand{\msun}{\ensuremath{{M}_\odot}}
\newcommand{\mch}{\ensuremath{M_\mathrm{Ch}}\xspace}
\def\lesssim{\mathrel{\hbox{\rlap{\hbox{\lower4pt\hbox{$\sim$}}}\hbox{$<$}}}}

\def\gtrsim{\mathrel{\hbox{\rlap{\hbox{\lower4pt\hbox{$\sim$}}}\hbox{$>$}}}}

\newcommand{\ie}{i.e.\xspace}

\def\apjl{ApJ}%
\def\apjs{ApJS}%
\def\ao{Appl.~Opt.}%
\def\aap{A\&A}%
\titlerunning{Nucleosynthesis imprints from Type Ia supernovae}
\authorrunning{F. Lach et al.}

\begin{document}

\title{Nucleosynthesis imprints from different Type Ia Supernova explosion scenarios and implications for galactic chemical evolution}

\author{F.~Lach\inst{1,2}\thanks{E-mail: florian.lach@h-its.org}
    \and F.~K.~Roepke\inst{1,3}
    \and I. R. Seitenzahl\inst{4}
    \and B. Cot\'{e}\inst{5,6,7}
    \and S.~Gronow\inst{1,2}
    \and A.~J.~Ruiter\inst{4}
}

\institute{ Heidelberger Institut f\"{u}r Theoretische Studien,
            Schloss-Wolfsbrunnenweg 35, D-69118 Heidelberg, Germany
  \and Zentrum f\"ur Astronomie der Universit\"at Heidelberg,
       Astronomisches Rechen-Institut, M\"{o}nchhofstr. 12-14, 69120 Heidelberg, Germany
  \and Zentrum f{\"u}r Astronomie der Universit{\"a}t Heidelberg,
       Institut f{\"u}r Theoretische Astrophysik, Philosophenweg 12,
       D-69120 Heidelberg, Germany
  \and School of Science, University of New South Wales, Australian Defence Force
       Academy, Canberra, ACT 2600, Australia
  \and Konkoly Observatory, Research Centre for Astronomy and Earth Sciences, MTA Centre for
       Excellence, Konkoly Thege Miklos 15-17, H-1121 Budapest, Hungary
       \and ELTE E\"{o}tv\"os Lor\'end University, Institute of Physics, Budapest, 1117, P\'azm\'any
       P\'eter S\'et\'any 1/A, Hungary
  \and National Superconducting Cyclotron Laboratory, Michigan State University, East Lansing, MI
       48824, USA
}

\date{23 June 2020 / 19 October 2020}

\abstract{
  We analyze the nucleosynthesis yields of various Type Ia supernova
  explosion simulations including pure detonations in
  sub-Chandrasekhar mass white dwarfs, double detonations and pure
  helium detonations of sub-Chandrasekhar mass white dwarfs with an
  accreted helium envelope, a violent merger model of two white dwarfs
  and deflagrations as well as delayed detonations in Chandrasekhar
  mass white dwarfs. We focus on the iron peak elements Mn, Zn and
  Cu. To this end, we also briefly review the different burning
  regimes and production sites of these elements as well as the
  results of abundance measurements and several galactic chemical
  evolution studies.

  We find that super-solar values of [Mn/Fe] are not restricted to
  Chandrasekhar mass explosion models. Scenarios including a helium
  detonation can significantly contribute to the production of Mn, in
  particular the models proposed for calcium-rich transients. Although
  Type Ia supernovae are often not accounted for as production sites
  of Zn and Cu, our models involving helium shell detonations can
  produce these elements in super-solar ratios relative to Fe.

  Our results suggest a re-consideration of Type Ia supernova yields
  in galactic chemical evolution models. A detailed comparison with
  observations can provide new insight into the progenitor and
  explosion channels of these events.}

\maketitle

\section{Introduction}
\label{sec:intro}

Over the past two decades, Type Ia supernovae (SNe~Ia) have been in
the focus of interest of astrophysical studies primarily because of
their application as distance indicators
\citep{riess1998a,perlmutter1999a} via the Phillips relation
\citep{phillips1993a}. As one of the signifcant sources for heavy
elements in the Universe, however, they are also important as a main
contributor to cosmic nucleosynthesis \citep{matteucci1986a,
  matteucci2001a, matteucci2006a, kobayashi2006a, kobayashi2009a,
  kobayashi2015a, kobayashi2019b}.  Although SNe~Ia have been
intensely studied in observational and theoretical approaches, the
questions concerning their progenitors and explosion mechanisms remain
open.

There is broad agreement that SNe~Ia originate from the thermonuclear
explosions of carbon-oxygen white dwarf (WD) stars
\citep{hoyle1960a}. In some cases, however, also oxygen-neon WDs
\citep{marquardt2015a} or hybrid carbon-oxygen-neon WDs
\citep{kromer2015a,willcox2016b} may give rise to similar events. In
most scenarios, the explosion is triggered by the interaction with a
binary companion.  This rather unsharp characterization leaves room
for a whole zoo of possible progenitor \citep[e.g.][]{wang2012b} and
explosion scenarios \citep[see, e.g.][]{hillebrandt2000a,
hillebrandt2013a}. Potential progenitors can be subdivided
into the single degenerate (SD) scenario involving one WD
\citep{whelan1973a} accompanied by a main sequence, giant, or helium
star and the double degenerate (DD) scenario consisting of a binary
system of two WDs \citep{iben1984a}. Another possibility, the
core-degenerate scenario, has been proposed by
\citet{kashi2011a}. Here a WD merges with a post-AGB core already
during the common envelope phase and forms a new WD above the
Chandrasekhar mass (\mch) which is stabilized by rotation. As the rotation
slows down it might explode as a SN~Ia.

From the explosion modeling point of view and the implied
nucleosynthesis output, however, the mass of the WD at the time of
explosion is the fundamental parameter
\citep{seitenzahl2017b}. Generally, one distinguishes near-\mch models
from models in which the exploding WDs are significantly below the
Chandrasekhar mass limit of approximately $1.4\,M_\odot$. Finally, the
characteristics of the explosion is governed by the combustion
mechanism. A thermonuclear combustion wave is formed via a runaway
process during convective burning, due to dynamical interaction in a
WD merger, or by converging shock waves. In the first case, a subsonic
deflagration propagates via heat conduction. The two other cases may
lead to the formation of a supersonic detonation where, in contrast,
the fuel is heated and burned by the compression of a shock wave (see,
e.g., \citealp{roepke2017a} for a review of thermonuclear combustion
in SNe~Ia).

It is not clear yet which of the various possible explosion mechanisms
can account for SNe~Ia. In addition to the bulk of normal SNe~Ia
obeying the Phillips relation a variety of subclasses of SNe~Ia have
been identified (see \citealp{taubenberger2017a} for a review) and
therefore it is most plausible that more than one scenario contributes
to the overall class of SNe~Ia.

One approach to check the validity of a certain scenario is to conduct
multidimensional hydrodynamical simulations of the explosion phase
together with the subsequent calculation of synthetic observables,
such as light curves and spectra.  These can then be compared to
observations of SNe~Ia and the initial model can be accepted,
discarded or adjusted accordingly. This exercise has been carried out
during the past years for a variety of explosion scenarios, and
although suitable explanations of sub-classes could be identified,
there is no fully convincing model for the bulk of normal SNe~Ia yet
\citep{blinnikov2006a, kasen2007b, blondin2011a,
  roepke2012a,seitenzahl2013a, sim2013a,
  fink2014a,kromer2013b,townsley2019double, gronow2020a}.

Another test of the realization of specific explosion scenarios are
abundance measurements in stars combined with galactic chemical
evolution (GCE) models \citep{matteucci2006a, travaglio2014a,
kobayashi2011b}.  Among other ingredients, these models assume certain
rates, delay-time distributions and nucleosynthesis yields for various
kinds of SNe~Ia and core-collapse supernovae (CCSNe).  The enrichment of
the investigated stellar population or galaxy with metals is then
compared to stellar abundances derived from spectroscopy and thus allows
to infer the origin of a particular element or group of elements. This
is the reason why the characteristic imprints of a certain explosion
scenario are of great interest.

A prominent example is the case of the element manganese.  It is widely
accepted that the primary contribution to Mn stems from SNe~Ia since the
observed values of [Mn/Fe] in the Galaxy increase from [Fe/H] $\approx
-1$ to the solar value and CCSN yields predict sub-solar values for
[Mn/Fe] \citep{timmes1995a, mcwilliam1997a, kobayashi2006a,
kobayashi2009a, kobayashi2011b, weinberg2019a, kobayashi2019b}.  This
``SN~Ia knee'' has already been explained by \citet{tinsley1979d},
\citet{greggio1983a}, and \citet{matteucci1986a} since it coincides
nicely with the decrease of the $\alpha$-elements, \ie [$\alpha$/Fe],
produced by CCSNe from their super-solar plateau at lower [Fe/H].
\citet{seitenzahl2013b} picked this up arguing that Mn is produced at
high burning densities in normal freeze-out from nuclear statistical
equilibrium (NSE, see Sect.~\ref{sec:models_burning}) and therefore it
predominantly originates from \mch explosions. The best agreement with
the data is achieved if SNe~Ia equally arise from sub-\mch and \mch
progenitors. These results have been further refined in the recent work
of \citet{eitner2020a}. Their non-LTE measurements of Mn in a sample of
42 stars in the Galaxy show a rather flat evolution of [Mn/Fe] near the
solar value lowering the contribution of \mch SNe~Ia to about 25\%.  The
trend in [Mn/Fe] is not so clear in dwarf spheroidal galaxies (dSphs)
but various works also find that a SN~Ia contribution to Mn is required
to explain observations \citep{cescutti2008a,north2012a,cescutti2017d}.
Another study carried out by \citet{mcwilliam2018a} claims that the most
metal-rich star in Ursa Minor, COS~171, was enriched by a low
metallicity, low-mass sub-\mch detonation. In particular its low [Mn/Fe]
and [Ni/Fe] values exclude a \mch origin. Furthermore, the sub-solar
amounts of Cu and Zn are also a hint for a low-mass progentior of the SN
explosion which has enriched the star COS 171 since these elements are
produced in strong $\alpha$-rich freeze-out.  Moreover, \citet{de2020a}
attribute different [Mn/Fe] values in dSphs to their specific star
formation history. The combination of sub-solar [Mn/Fe] and a short star
burst, as seen in Sculptor, indicates a dominant role of sub-\mch
explosions. In contrast, Fornax and Leo I show a long-lasting star
formation and [Mn/Fe] around the solar value which points to an
increasing enrichment via near-\mch SNe~Ia.

Our work analyzes the nucleosynthetic yields of various models for SNe
Ia, namely pure deflagrations and delayed detonations in $M_\mathrm{Ch}$
WDs as well as pure detonations and double detonations of
sub-$M_\mathrm{Ch}$ WDs. The aim is not to go into detail about
observational implications of specific isotopes and to evaluate whether
the model matches any observed SN~Ia but to identify specific abundance
patterns that are characteristic for particular explosion scenarios and
therefore to constrain whether this scenario is required to produce the
observed elemental and isotopic abundances. The key discriminant is the
initial mass of the WD that sets its central density, mainly determining
the freeze-out regime reached by the burning region. So-called normal
freeze-out from NSE, for instance, requires high densities only reached
in $M_\mathrm{Ch}$ WDs, and the presence of a helium detonation yields
unique abundance patterns not produced in explosive carbon-oxygen
burning.  Moreover, electron captures significantly reduce the electron
fraction at high densities and therefore shift the nucleosynthesis
yields in NSE to more neutron rich isotopes (see
Sect.~\ref{subsec:electroncap}). Thus, the occurrence of iron group
elements (IGEs) with a considerable neutron excess is a hint towards the
Chandrasekhar mass scenario \citep{yamaguchi2015a}.

This work is structured in the following way: In
Sect.~\ref{sec:models_burning} we first review the different burning
regimes and then summarize the explosion models investigated.
Section~\ref{sec:yields} presents the results of our nucleosynthesis
analysis and we discuss implications of these results for the nature of
the progenitor for SNe~Ia. The elements Mn, Zn and Cu are of particular
interest. Because we find substantial amounts of the unstable
radionuclides $^{68}$Ge, $^{68}$Ga, and $^{65}$Zn, we briefly test their
potential impact on the observables. In Sect.~\ref{sec:summary} we
summarize our findings.

\section{Explosion models and burning regimes}
\label{sec:models_burning}

\subsection{Explosive silicon burning} 
\label{subsec:explsiliburn}

\begin{figure}[htbp]
  \centering
  \def\svgwidth{256pt}
  \tiny
  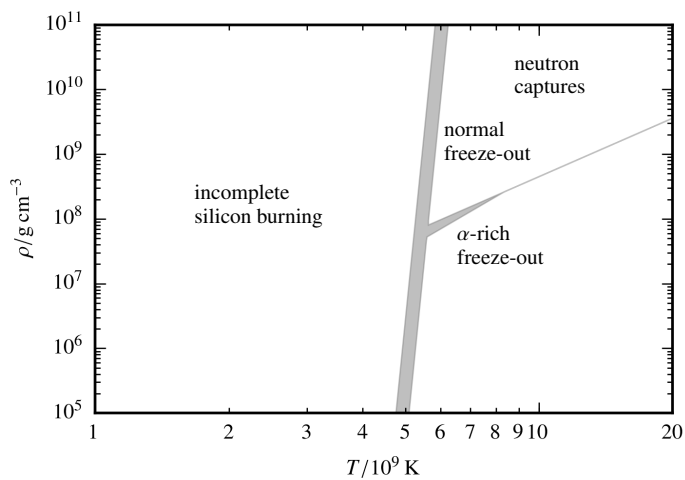
  \caption{Burning regimes in the $T-\rho$-plane according to
    \citet{woosley1973a}. Below a certain temperature the burning of
    silicon is incomplete. At high temperatures the state of nuclear
    statistical equilibrium is achieved and silicon is transformed into
    iron peak elements. This region is subdivided in the regime of
    normal freeze-out and $\alpha$-rich freeze-out. The shaded area
    covers values of $\chi$ between $1$ and $10$ according to
    Eqs.~(\ref{eq:si_exh}) and (\ref{eq:freezeout}).}
  \label{fig:freezeregimes}
\end{figure}

In a thermonuclear supernova explosion the innermost, \ie densest,
part of the WD star reaches temperatures sufficiently high for silicon
burning and thus significantly contributes to the production of
IGEs. \citet{woosley1973a} divide the parameter space of this burning
regime into three regions: the regime of incomplete silicon burning,
alpha-rich freeze-out, and normal freeze-out from NSE (see
Fig.~\ref{fig:freezeregimes}). In NSE all the abundances from protons,
neutrons and $\alpha$-particles up to the iron peak have reached an
equilibrium, \ie forward and reverse reactions cancel each
other. Besides the peak temperature $T_\mathrm{peak}$ and density
$\rho_\mathrm{peak}$ also the time scale on which a particular fluid
element cools after the crossing of the burning front determines the
nucleosynthesis yields. Adiabatic cooling can be written as
\citep{woosley1973a}
\begin{align} 
  \rho(t)&= \rho_{\text{peak}}
  e^{-t/\tau_{HD}} \label{rhotimescale}, \\ T_9(t)&= T_{9,\text{peak}}
  \left(\frac{\rho(t)}{\rho_{\text{peak}}}\right)^{\gamma-1}.
  \label{Ttimescale} 
\end{align} 
Here $\gamma$ denotes the adiabatic exponent and $\tau_{HD} = 446 \chi
\rho_{\text{peak}}^{-1/2} \ $ is the hydrodynamical time scale. With the
help of the scaling parameter $\chi$, the dependency of the results on
the time scale can be examined.

\citet{meakin2009a} present an updated prescription for the density
evolution in SNe~Ia. They employ an exponential temperature decay and
ensure adiabaticity by fixing the entropy to the post-burning state. The
entropy $S$ is a function of temperature, density and averaged values of
the mass number $\bar{A}$ and the proton number $\bar{Z}$. Thus, the
density can be obtained from $S=S(T(t),\rho(t),\bar{A},\bar{Z})$.
However, for demonstration purposes we stick to the formulation of
\citet{woosley1973a} in the following since the main statements about
freeze-out from NSE are not affected by the exact choice of the
expansion prescription.

Incomplete silicon burning is characterized by the presence of two
quasi-equilibrium clusters, \ie  only certain regions in the chart of
nuclei that have reached equilibrium states, centered around $^{28}$Si
and $^{56}$Ni, respectively. These are separated by the bottleneck at a
mass number of $A \approx 45$ (e.g. Ti and Sc). These elements are
weakly bound compared to Ca and are therefore low in abundance. Hence,
the flow of material through the bottleneck is very low and intermediate
mass elements (IMEs) as well as IGEs remain present after the burning is
quenched. At higher temperatures the bottleneck is removed. The
equilibrium clusters merge and matter achieves the state of NSE,
ultimately converting silicon to IGEs almost completely. The temperature
necessary for silicon exhaustion, that is $X(^{28}$Si$) \lesssim 5
\times 10^{-3}$, can be approximated by
\citep{woosley1973a}
\begin{align}
  T_{9,\text{peak}} \gtrsim 4.3
  \left(\frac{\rho_{\text{peak}}}{\chi^{2}}\right)^{1/68}.  
  \label{eq:si_exh}
\end{align}
In NSE, the abundance distribution is uniquely determined by density,
temperature, and electron fraction $Y_e$. The composition changes as
soon as the first reactions drop out of equilibrium due to decreasing
temperature. At high densities, matter is characterized by a low
fraction of light, free particles such as neutrons, protons and
$\alpha$-particles. Therefore, the composition during this
\textit{normal freeze-out} or \textit{particle-poor freeze-out} is not
altered significantly by the capture of light particles during
expansion. At lower densities, in contrast, light particles are more
abundant and thus react with the prevailing iron group nuclei and
bring matter out of NSE composition. Due to the high mass fraction of
$\alpha$-particles this drop-out of equilibrium is called
\textit{alpha-rich freeze-out}.  \citet{woosley1973a} derive the
approximate relation
\begin{align}
  \rho_\mathrm{peak} \lesssim \min
\begin{cases} 4.5 \times 10^5T_{9,\text{peak}}^3 \\ 2.5 \times
    10^5T_{9,\text{peak}}^4 \chi^{-2/3}  \end{cases} ,
  \label{eq:freezeout}
\end{align}
for the density separating $\alpha$-rich freeze-out and normal
freeze-out using $\gamma=4/3$ in Eq.~(\ref{rhotimescale}) and
Eq.~(\ref{Ttimescale}).

Therefore, the $T_\mathrm{peak}-\rho_\mathrm{peak}$-plane is split
into three regions by Eqs.~(\ref{eq:si_exh}) and (\ref{eq:freezeout})
as illustrated in Fig.~\ref{fig:freezeregimes} (see also Fig.~20 of
\citealp{woosley1973a}). These regions produce different chemical
compositions after the last reactions have frozen out. This is also
visualized in Fig.~\ref{fig:ni56} for the different explosion models
explained in the next Section. The separation of the different regions
can be shifted continuously by a variation of the scaling parameter
$\chi$.  The gray shaded area in Figs.~\ref{fig:freezeregimes} and
\ref{fig:ni56} covers values of $\chi$ between $1$ and $10$.

\subsection{Neutronization in high density material}
\label{subsec:electroncap}

We have discussed in the previous Section how the NSE composition is
altered in the two different freeze-out regimes but the NSE
composition itself is determined by the electron fraction $Y_e$
(related to the neutron excess $\eta = 1-2Y_e$). In NSE, but also in
quasi-statistical equilibrium (QSE), the most abundant nuclei after
freeze-out are those with the highest binding energy and an electron
fraction close to that of the initial fuel. For symmetric matter,
\ie $Y_e=0.5$, the most abundant nucleus is $^{56}$Ni. As $Y_e$
decreases isotopes like $^{57-60}$Ni and $^{54-58}$Fe become more
abundant depending on the actual electron fraction and their binding
energies. The value of $Y_e$ is determined and altered by three
mechanisms.

First, the electron fraction of the WD is set by the metallicity $Z$
of the progenitor main sequence star. Most important are the
abundances of $^{56}$Fe and CNO nuclei which are mostly converted to
$^{14}$N in hydrogen burning (CNO-cylce) and subsequently to $^{22}$Ne
in helium burning. These two isotopes, $^{56}$Fe and $^{22}$Ne,
provide the dominant part of the surplus of neutrons in the exploding
WD. \citet{timmes2003a} derive an approximate relation for the mass of
$^{56}$Ni as a function of metallicity. It shows a decreasing trend
for $^{56}$Ni with increasing $Z$ and a variation of 25\% if Z is
varied by a factor of three.

Second, the initial $Y_e$ of the WD is altered during convective
carbon burning (``simmering'') preceding the thermonuclear runaway in
a \mch WD. The number of neutrons is increased via the capture of free
electrons onto the highly ionized atoms \citep{bahcall1964a}:
\begin{align}
  e^- + (Z,A) \to (Z-1,A) + \nu.
  \label{eq:e-capture}
\end{align}
These endoergic reactions require high electron energies and therefore
become important at high densities only. \citet{chamulak2008a} study the
behavior of $Y_e$ during carbon burning and find that the reaction chain
$^{12}\mathrm{C}(p,\gamma)^{13}\mathrm{N}(e^-,\nu)^{13}\mathrm{C}$ is
the dominant mechanism for reducing $Y_e$ for densities around $1\times
10^9\,\si{g.cm^{-3}}$. The electron capture on $^{13}$N is gradually
replaced by
$^{12}\mathrm{C}(^{12}\mathrm{C},p)^{23}\mathrm{Na}(e^-,\nu)^{23}\mathrm{Ne}$
for $\rho > 1.7\times 10^9\,\si{g.cm^{-3}}$. \citet{chamulak2008a}
estimate a maximal reduction in $Y_e$ by $6.3 \times 10^{-4}$ and
\citet{piro2008c} suggest a carbon-dependent value of $|\Delta
Y_{e,\mathrm{max}}|=1.7 \times 10^{-3} \, X(^{12}\mathrm{C})$.

Finally, the most dramatic changes to the electron fraction happen
during the explosion itself.  In NSE, electrons are mostly captured by
free protons at high temperatures. Subsequently, as the temperature
drops IGEs capture most of the electrons. This leads to a decrease in
$Y_e$ down to $Y_e \approx 0.44$ in the most extreme cases in the
central regions of the exploding WD star. Therefore, the regions with
the highest density do not contribute significantly to the production of
$^{56}$Ni \citep{brachwitz2000a} but to more neutron-rich IGEs. In
general the nucleosynthesis results then depend on the central density
and the corresponding density gradient. Detailed studies of the
nucleosynthesis in Chandrasekhar-mass models and the effect of
neutronization have been carried out by \citet{thielemann1986a},
\citet{iwamoto1999a}, and, with updated electron capture rates
\citep{langanke2000a}, by \citet{brachwitz2000a} and \citet{bravo2019b}.
When 3D effects in deflagration models are taken into account the final
abundance stratification will be smoothed out compared to the
theoretical prediction based on density, metallicity and neutronization
\citep{seitenzahl2013a}.

In summary, the neutronization due to electron captures is an effect
restricted to the high densities reached only in WDs close to \mch and
hence the abundances of very neutron-rich isotopes are a hint to the
Chandrasekhar-mass scenario. The abundance ratio of nickel to iron, for
instance, measured in late time spectra of SNe~Ia can be taken as a
proxy for the quotient $^{58}$Ni$/^{56}$Ni and thus for the
neutronization. This has been used by \citet{flors2019a} to infer the
contribution of sub-\mch progenitors to SNe~Ia in relation to \mch
progenitors .

\subsection{Explosive helium burning}
\label{subsec:explosivehelium}

In addition to explosive silicon burning, the burning of helium is
another source of nucleosynthesis products in the double detonation
scenario (see Sect.~\ref{sec:intro}). Explosive helium burning has been
studied by \citet{khokhlov1984a} and \citet{khokhlov1985a}. They found
that in general the burning is characterized by a competition between
the triple-$\alpha$ reaction and $\alpha$-captures on heavier nuclei.
First, $^{12}$C is synthesized by the reaction $3\alpha \to ^{12}$C and
subsequently heavier $\alpha$-elements ($^{16}$O, $^{20}$Ne, $^{24}$Mg,
$^{28}$Si, $^{32}$S, $^{36}$Ar, $^{40}$Ca, $^{44}$Ti, \ldots) are
produced. The time scale for the capture of an $\alpha$-particle
increases for higher mass numbers $A$ due to the higher Coulomb
barriers. The latter is penetrated more easily at higher particle
energies and the time scale therefore is  temperature-dependent.
Consequently, the $\alpha$-chain stops as soon as the time scale for the
3$\alpha$-reaction is shorter than for the next $\alpha$-capture.
However, above a certain, density-dependent temperature of approximately
$2\times 10^{9}\,$K at $\rho = 5\times 10 ^{6}\,\si{g.cm^{-3}}$,
$^{56}$Ni is always the most abundant isotope since it has the highest
binding energy at $Y_e = 0.5$ (see Sect.~\ref{subsec:explsiliburn}).
These temperatures are usually surpassed in helium detonations, and
therefore most of the material is converted to $^{56}$Ni. Nevertheless,
there is a way to stop the $\alpha$-chain before $^{56}$Ni is reached
also at high temperatures: If the initial fuel is polluted with carbon,
oxygen or nitrogen, for instance, the slow triple-$\alpha$ reaction is
bypassed by $\alpha$-captures on these seed nuclei. This leads to a very
fast depletion of $\alpha$-particles and the nucleosynthesis stops below
$A=56$ once the material runs out of $\alpha$-particles (see also
\citealp{woosley2011b}, \citealp{shen2014b, gronow2020a}).

Beyond $A=56$, the reverse reactions become increasingly important and
start to balance the $\alpha$-captures to some extent. Nevertheless, a
high abundance of $\alpha$-particles results in an enhanced production
of elements beyond Ni such as Cu and Zn compared to their NSE abundance.
The time scale to reach NSE is about $1\,\mathrm{s}$ at a temperature of
$5\times 10^{9}\,$K. However, conditions necessary for NSE are not
achieved in the major part of the helium detonation and thus the
nucleosynthesis exhibits interesting differences to the burning products
of the CO core.
\begin{figure*}
  \centering
  \includegraphics[width=\textwidth]{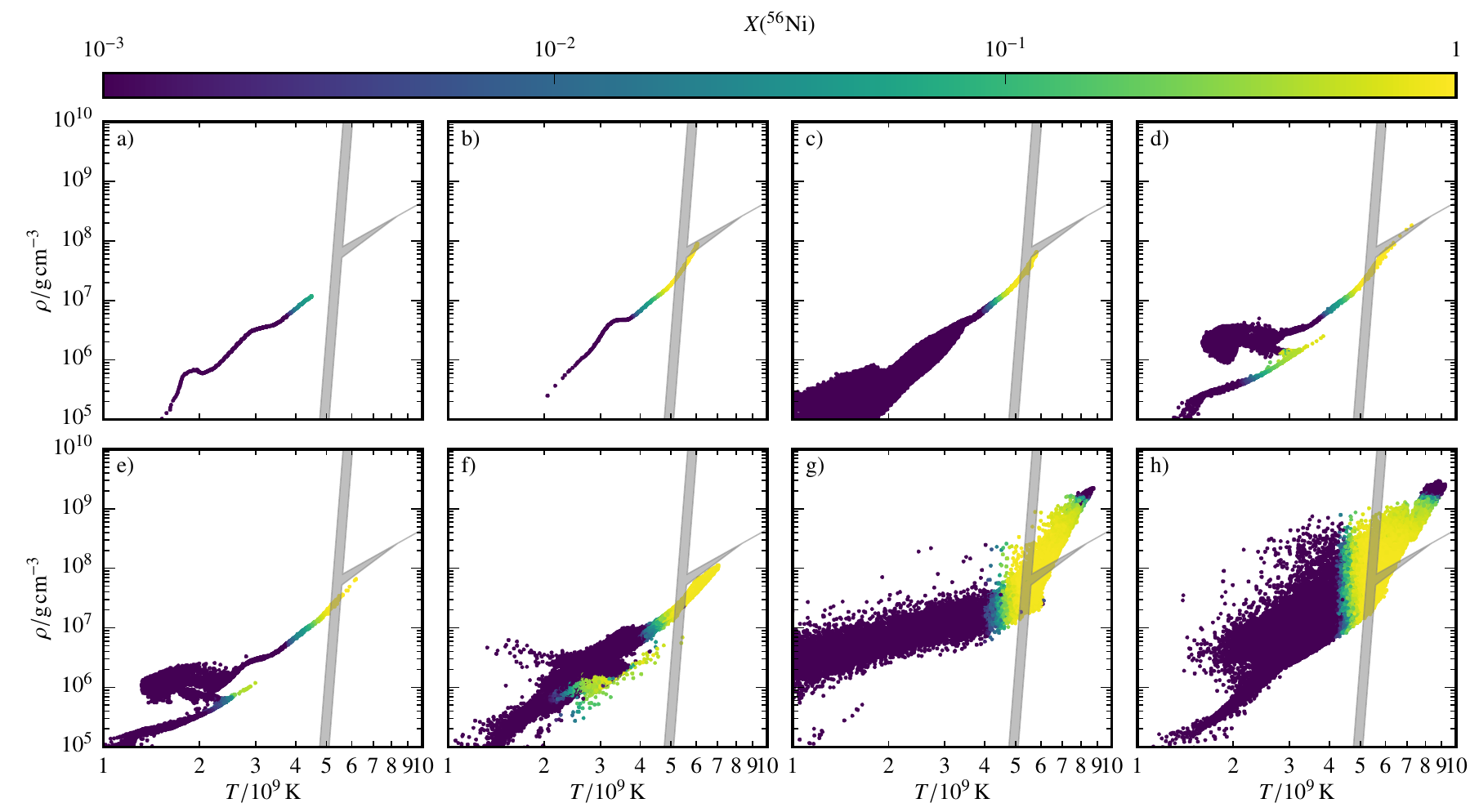}
  \caption{Distribution of the tracer particles in the
    $T_\mathrm{peak}-\rho_\mathrm{peak}$~-~plane with color coded mass
    fraction of $^{56}$Ni $100\,\si{s}$ after triggering the explosion. The 
    panels show: a) PD081, b) PD115, c) VM, d) CSDD-S, e) CSDD-L, f)
    M2a$_\odot$, g) R60, h) N100ddt. The grey shaded regions separate
    incomplete (left) from complete silicon burning (right) and normal
    freeze-out (upper right) from $\alpha$-rich freeze-out (lower
    right) according to Eqs.~(\ref{eq:si_exh}) and
    (\ref{eq:freezeout}). The area covered corresponds to a varying
    scaling parameter: $1<\chi<10$.}
  \label{fig:ni56}
\end{figure*}

\subsection{Models}
\label{subsec:models}

In this work, we investigate three distinct kinds of explosion
models. The first class consists of pure detonations of
sub-Chandrasekhar mass CO WDs. We closely examine the violent merger
(VM) of a $0.9\,M_\odot$ with a $1.1\,M_\odot$ WD simulated by

\citet{pakmor2012a}. In addition, two pure detonations of CO WDs with
total masses of the progenitor of $0.81$ (PD081) and $1.15\,M_\odot$
(PD115) of \citet{sim2010a} are included in our analysis.

Secondly, we study models including a detonation of a helium shell on
top of a sub-\mch mass WD (``.Ia''-SN,
\citealp{bildsten2007a,shen2009a,shen2010a}) eventually triggering a
second detonation burning the CO core (double detonation).  Two models
are taken from \citet{sim2012a} which follow the explosion of low-mass
CO cores ($0.58$ and $0.45\,M_\odot$) with a prominent helium shell of
$0.21\,M_\odot$ (hereafter CSDD-S and CSDD-L, respectively). The second
detonation is triggered via the converging shock mechanism not far from
the center of the WD \citep{fink2007a}. The HeD-S model follows the same
setup as CSDD-S, but the detonation of the core is suppressed. These
pure helium detonations are a possible explosion mechanism for Ca-rich
transients \citep[e.g.][]{inserra2015a}.  Moreover, a double detonation
in the core of a $1.05\,M_\odot$ WD with a carbon-enriched helium shell
of $0.073\,M_\odot$ (Model M2a of \citealp{gronow2020a}) is included. In
this case, a detonation in the core is triggered at its outer edge when
the helium detonation front converges on the far side of its ignition.
In addition, an equivalent model to M2a at solar metallicity M2a$_\odot$
is investigated.  This model is set up as a WD with a total mass of
$1.06\,M_\odot$. Its core consists of $^{12}$C, $^{16}$O, and -- in
order to reproduce the metallicity -- 1.34\% of $^{22}$Ne. In the
$^{4}$He shell ($M_\mathrm{shell} = 0.075\,M_\odot$) an admixture of
0.34\% by mass of $^{14}$N accounts for solar metallicity in for the
hydrodynamical explosion simulation.

Finally, the $M_\mathrm{Ch}$ scenario is analyzed. We add the N5def
model of \citet{fink2014a} as an example of a pure deflagration, the
N100ddt delayed detonation identical with model N100 of
\citet{seitenzahl2013a} as well as a pure deflagration in a \mch WD with
central density of $2.6\times10^{9}\si{g.cm^{-3}}$.  The latter (R60) is
ignited in a single spot $60\, \mathrm{km}$ off-center and, like N5def,
does not disrupt the whole star but leaves behind a bound remnant.  The
model produces a very faint explosion ejecting only $0.018\,\msun$ of
$^{56}$Ni and  $0.049\,\msun$ of material in total. Moreover, the
kinetic energy of the ejecta amounts to $8.17\times 10^{48}\si{erg}$ and
an intact WD of $1.33\,\msun$ is left behind after the explosion. This
model has been calculated for this paper with methods similar to
\citet{fink2014a} but with an updated equation of state
(\citealp{timmes1999a}) and a gravity solver based on fast Fourier
transforms.

More details concerning the individual setups and the employed codes can
be found in the references above.  Moreover, data for the models VM,
CSDD-S, CSDD-L, HeD-S, N5def, N100ddt and M2a have been made publicly
available in the online model database \textsc{HESMA}
\citep{kromer2017a}. An important fact to be aware of is that all
hydrodynamic explosion simulations were done using the \textsc{LEAFS}
code \citep{reinecke1999b,reinecke2002b,roepke2005c} except for the
double detonations M2a and M2a$_\odot$ which were computed with
\textsc{arepo} \citep{springel2010a}.

Because solving a large nuclear network in parallel with the actual
explosion simulation goes beyond the scope of the current computational
resources the nucleosynthesis yields are calculated in a post-processing
step.  To this end, virtual tracer particles are placed into the
exploding WD star and advected passively with the fluid flow recording
their thermodynamic history. Subsequently, this data is used to
determine the isotopic abundances produced by the explosion.

Because we partially work with models from \textsc{HESMA} but also add
new explosion simulations, the postprocessing and the treatment of
metallicity in it follows different approaches: Models CSDD-L, CSDD-S,
HeD-S, PD081, PD115 were post-processed at zero metallicity and VM,
N100ddt, and N5def at solar metallicity using the methods described in
\citet{travaglio2004a}. In the latter models, solar metallicity was
mimicked by including 2.5\% of $^{22}$Ne to approximately adjust $Y_e$
to the solar value (see also \citealp{seitenzahl2013a}). All other
simulations, i.e.\ M2a, M2a$_\odot$, and R60,  use the \textsc{YANN}
code \citep{pakmor2012b} for postprocessing. The metallicity of the
material in M2a is assumed to be zero while M2a$_\odot$ and R60 apply
solar metallicity. To this end, the abundances of all isotopes in the
core (``carbon-oxygen'') material are set according to
\citet{asplund2009a} except for elements lighter than fluorine. While H
and He are ignored, all CNO isotopes are converted to $^{22}$Ne thus
accounting for their processing in H and He burning. In the shell
material of M2a$_\odot$ carbon and oxygen are instead converted to
$^{14}$N. To check the effect of the different metallicity
implementations we also post-processed R60 and M2a$_\odot$ with the
appropriate amounts of $^{22}$Ne in CO material and $^{14}$N in the
helium shell, respectively, but no other isotopes present. These
variants are labeled R60$_\mathrm{Ne}$ and M2a$_\mathrm{Ne}$. In all
cases the 384 species network of \citet{travaglio2004a} is utilized.
Reaction rates were taken from the \textsc{REACLIB} database
(\citealp{rauscher2000a}, updated 2009) and only for our most recent
simulations (R60, M2a etc.) the version of 2014 is applied.

\section{Discussion of the nucleosynthesis yields}
\label{sec:yields}

\begin{figure*}[htbp]
  \centering
  \includegraphics[width=\textwidth]{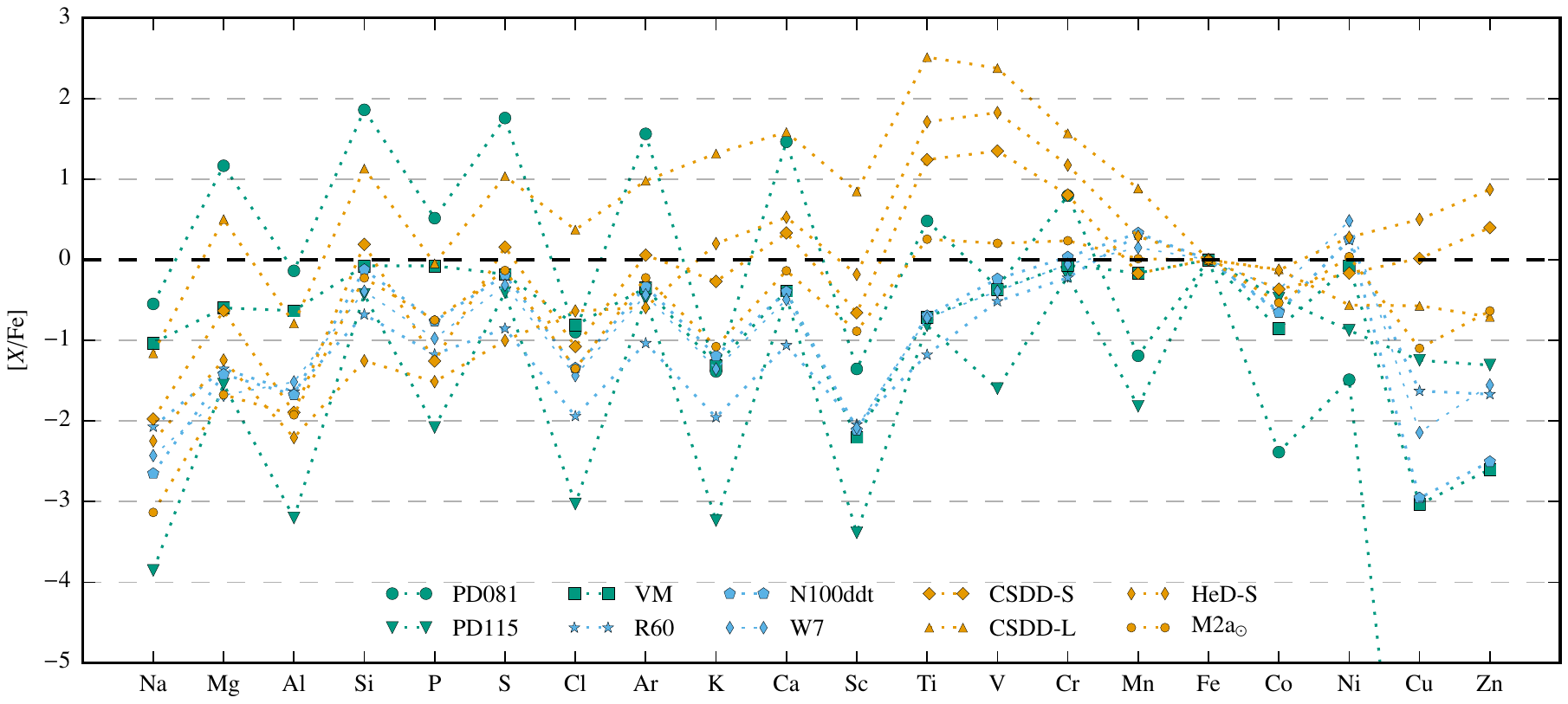}
  \caption{Elemental ratios to Fe (with radioacitve isotopes
    decayed to $2 \times 10^9\,\si{yr}$) compared to their solar ratios
    for three sub-\mch detonations (VM, PD081, PD115), three double
    detonations (M2a$_\odot$, CSDD-S, CSDD-L), one helium detonation
    (HeD-S), two pure $M_\mathrm{Ch}$ deflagrations (R60, W7) and a
  delayed detonation (N100ddt).  }
  \label{fig:ratio_fe}
\end{figure*}

An instructive way to illustrate the burning regimes reached by a
specific model is the distribution of the tracer particles in the
$T_\mathrm{peak}-\rho_\mathrm{peak}$-diagram shown in
Fig.~\ref{fig:freezeregimes}.

In the case of the sub-Chandrasekhar mass pure detonations (VM, PD081,
PD115) the important characteristic from the nucleosynthesis point of
view is the low central density ($\rho_c \lesssim
10^8\,\si{g.cm^{-3}}$) of the WD. Therefore, IGEs are produced in
alpha-rich freeze-out and incomplete silicon burning only (see
Fig.~\ref{fig:ni56}).  While the higher-mass CO core (PD115) produces
$^{56}$Ni also in NSE ($\alpha$-rich freeze-out) the low-mass core
model (PD081) synthesizes all of its $^{56}$Ni in incomplete silicon
burning. Also the fuel in the VM model stays below densities of $8
\times 10^7\,\si{g.cm^{-3}}$ and therefore it burns to NSE in the
$\alpha$-rich freeze-out regime. Although in the merger process the
structure of the primary is not much affected, the disruption of the
secondary causes the larger scatter in the thermodynamic properties of
the tracer particles at lower densities.

In addition to this, the double detonation models (CSDD-S, CSDD-L,
M2a$_\odot$) clearly show that $^{56}$Ni is also produced in the
helium shell detonation. The tracer particles located in the shell
cover an area in the parameter space shown in Fig.~\ref{fig:ni56} that
is slightly below that of the core detonation. The scatter in the \mch
deflagration models is larger due to the turbulent motion of the flame
and the pre-expansion of the WD during the burning phase. This is even
more obvious in the delayed detonation model since some tracers might
be affected first by the deflagration and subsequently by the
detonation. We note that for the single-spot ignited Model R60 the WD is
not disrupted completely and hence Fig.~\ref{fig:ni56} only shows
ejected tracers. Because only such massive WDs reach high densities of
$\rho \gtrsim 10^9\,\si{g.cm^{-3}}$ and consequently their
nucleosynthesis yields originate mainly from normal freeze-out, these
kinds of explosions contribute to elements not synthesized in sub-\mch
models.

The gross nucleosynthesis yields are summarized in
Fig.~\ref{fig:ratio_fe}, which displays the elemental ratio to iron
compared to the solar ratio according to \citet{asplund2009a}. Since
the PD081 model does not burn to NSE, it synthesizes super-solar
amounts of IMEs showing a strong odd-even effect. Furthermore, it
exhibits a rather high Cr abundance and drops off steeply for higher
mass IGEs. CSDD-S also produces super-solar abundances of some IMEs,
such as Si, S, Ar and Ca which is not surprising as it shares the
characteristics of a low-mass CO core with the PD081 model. However,
the helium shell detonation adds substantial amounts of light IGEs
(Ti, V, Cr) as well as Cu and Zn to the mix. These elements are most
abundant in HeD-S since they are primarily produced in the helium
detonation. Moreover, it should be noted that HeD-S as well as CSDD-L
yield super-solar amounts of Mn. The VM yields behave rather
inconspicuous not showing any overproduction but instead significant
underproductions in Mn, Co, Cu and Zn. Interestingly, the M2a$_\odot$
simulation exhibits characteristics similar to those seen in pure
detonations (a strong odd-even effect for IMEs, an underproduction of
Co, a drop-off for Cu and Zn) as well as helium shell detonation
features (super-solar values of Ti, V, Cr and solar abundance of
Mn). Finally, the pure deflagration explosions R60 and W7 (we included
the W7 model from \citet{iwamoto1999a} because it has been widely used
in GCE calculations) and the delayed detonation N100ddt display low
abundances of the light IGEs V and Cr and super-solar abundances of Mn
and stable Ni. Moreover, the produced amounts of Cu and Zn are
negligible as is also the case for the pure detonations VM and PD081.

We summarize that all SNe~Ia models included here underproduce Co
compared to the solar value.  Moreover, we confirm the known fact that
$M_\mathrm{Ch}$ explosions can produce Mn at super-solar values and
that they also overproduce stable Ni. This fact has been used to
discriminate $M_\mathrm{Ch}$ from sub-$M_\mathrm{Ch}$ explosions in
nebular spectra \citep{flors2019a}. The theoretical reason for this
distinction is that stable Ni ($^{58,60}$Ni) is produced at higher
densities due to the lower electron fraction (see
Sect.~\ref{subsec:electroncap}). However, we also find the double
detonation models can overproduce Mn as well as Ni, which makes the
nucleosynthetic distinction between \mch and sub-\mch explosions not
as straightforward as expected. The lighter elements Ti, V and Cr stem
either from He detonations or from the incomplete silicon burning
region in low-mass CO cores.  The heaviest IGEs Cu and Zn are
synthesized in He detonations only and therefore test the double
detonation scenario: any observations and GCE calculations finding Cu
or Zn to be produced by SNe~Ia would hint to the occurrence of this
explosion mechanism.

\subsection{Manganese}
\label{subsec:manganese}

\begin{figure*}[htbp]
    \centering
    \includegraphics[width=\textwidth]{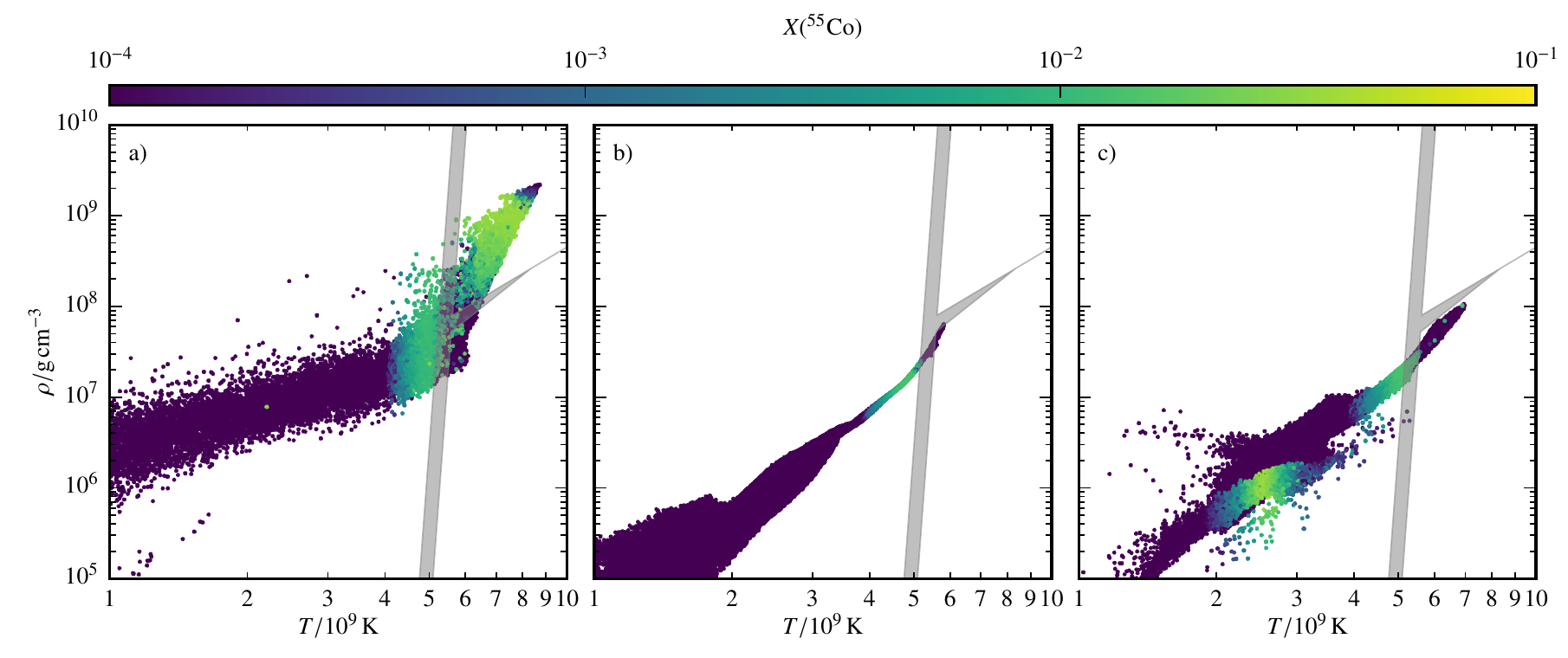}
    \caption{Same as Fig.~\ref{fig:ni56} with a color coded mass
      fraction of $^{55}$Co. The panels show: a) R60, b) VM, c)
      M2a$_\odot$}
    \label{fig:co55}
  \end{figure*}

\begin{table*}[h]
  \centering
  \caption{Mn to Fe ratio compared to solar for the total explosion
    at 100$\,$s after ignition as well as shell and core
    material only, the total amount of stable Mn in $M_\odot$ and the
    radioactive and stable isotopes from which
    $^{55}$Mn originates (in percent).}
  \begin{tabularx}{0.8\textwidth}{c @{\extracolsep{\fill}}cccccccc}
    \toprule
    model  &  [Mn/Fe] & [Mn/Fe]$_\mathrm{shell}$ & [Mn/Fe]$_\mathrm{core}$ & Mn  &  $^{55}$Co  & $^{55}$Fe & $^{55}$Mn  \\
    \midrule
    PD081  & -1.19 & -  & -  & 4.90e-06& 99.3 & 0.7 & -     \\ 
    PD115  & -1.82 & -  & -  & 1.04e-06& 97.6 & 2.4 & -     \\ 
    VM     & -0.16 & -  & -  & 3.74e-03& 97.9 & 2.1 & -     \\ 
    M2a$_\odot$ &  -0.03 & 0.81 & -0.07 & 4.57e-03 & 98.9 & 1.1 & -     \\
    M2a$_{\mathrm{Ne}}$ & -0.15 & 0.81 & -0.21 & 3.46e-03 & 99.5 & 0.5  & -  \\
    M2a    & -1.00 & 0.53 & -1.63 & 4.94e-04& 99.7 & 0.3 & -     \\
    CSDD-L &  0.88 & 1.69 & -1.19 & 1.47e-03& 99.6 & 0.4 & -     \\
    CSDD-S & -0.17 & 0.29 & -1.37 & 1.27e-03& 99.9 & 0.1 & -     \\
    HeD-S  &  0.29 & 0.29 &  -    & 1.21e-03& 99.9 & 0.1 & -     \\
    N5def  &  0.35 & -  & -  & 4.03e-03& 87.5 & 12.3 & 0.2  \\ 
    N100ddt&  0.33 & -  & -  & 1.33e-02& 85.8 & 14.0 & 0.2  \\ 
    R60    &  0.11 & -  & -  &
    2.30e-04 & 85.0 & 14.8  & 0.2  \\
    R60$_{\mathrm{Ne}}$    &  0.11 & -  & -  &
    2.30e-04 & 85.1 & 14.7  & 0.2  \\
    W7     &  0.16 & -  & -  & 8.87e-03&  -  &  -  &  -     \\ 
    \bottomrule
  \end{tabularx}
  \label{tab:manganese}
\end{table*}

The only stable isotope of Mn, $^{55}$Mn, is produced in CCSNe as well
as in SNe~Ia in incomplete silicon burning primarily via the channel
$^{55}$Co$\to ^{55}$Fe$\to ^{55}$Mn. As already pointed out by
\citet{seitenzahl2013b}, a super-solar production of Mn is required to
explain the rise in [Mn/Fe] for $\mathrm{[Fe/H]} \gtrsim -1$ to its
solar value. Although the CCSN contribution to Mn is uncertain, all
current models predict [Mn/Fe] ratios below the solar value. Therefore,
$M_\mathrm{Ch}$ explosions must be added to the mix of SNe~Ia. Only
these WDs reach densities high enough for normal freeze-out and thus
offer an additional site of production to the regime of incomplete
silicon burning. The production of $^{55}$Co is illustrated in
Fig.~\ref{fig:co55} for the pure deflagration model R60, the violent
merger model VM and the double detonation model M2a$_\odot$. It clearly
shows that $^{55}$Co is synthesized in normal freeze-out in the R60
model in contrast to VM and M2a$_\odot$. Furthermore, it can be seen
that Mn also originates from the He detonation in Model M2a$_\odot$.
Table~\ref{tab:manganese} summarizes the Mn yields and also gives the
fraction of the isotopes contributing to the final Mn abundance. It
reveals that only $M_\mathrm{Ch}$ deflagrations (N5def, N100ddt, R60,
R60$_\mathrm{Ne}$, W7), the double detonation CSDD-L, and the helium
detonation HeD-S achieve super-solar [Mn/Fe] ratios. For model R60  the
[Mn/Fe] values are somewhat lower compared to N100ddt and N5def due to
the newer set of reaction rates employed in the postprocessing. While
M2a$_\odot$ roughly reaches the solar value of [Mn/Fe] the other double
detonations, i.e. M2a, M2a$_\mathrm{Ne}$ and CSDD-S, exhibit a sub-solar
production of [Mn/Fe]. 

The value of [Mn/Fe] in the double detonation models is governed by
three fundamental parameters of the initial model:
\begin{enumerate}[(i)]
  \item The ratio of shell to core mass
    ($M_\mathrm{shell}/M_\mathrm{core}$) is a crucial factor. Most Mn
    is produced in the helium detonation which is reflected by the
    super-solar value of 0.29 for HeD-S (see Table~\ref{tab:manganese}). In
    contrast to that [Mn/Fe] is below solar in CSDD-S due to the
    sub-solar value of the core detonation.  The WD in Model CSDD-L,
    however, has a very low core mass ($0.45\,M_\odot$) and a helium
    shell of $0.2\,M_\odot$ (the same as CSDD-S) and therefore the
    contribution of the core detonation to [Mn/Fe] is less
    significant.
  \item The density of the helium shell also affects [Mn/Fe]. $^{55}$Co
    is produced in a rather well-defined range of initial densities in
    the helium envelope above approximately $6 \times
    10^{5}\,\si{g.cm^{-3}}$. The density at the base of the helium shell
    for the low-mass progenitors in CSDD-L ($\rho_\mathrm{c} =
    5.92\times 10^{5}\,\si{g.cm^{-3}}$) as well as CSDD-S in
    ($\rho_\mathrm{c}=12 \times 10^{5}\,\si{g.cm^{-3}}$) is sufficient
    to synthesize $^{55}$Co. Thus, the amount of Mn is quite similar in
    both models. However, the amount of Fe is much lower in CSDD-L due
    to the lower density in the envelope and therefore the discrepancy
    between CSDD-L and CSDD-S in [Mn/Fe] mainly originates from the
    yields of the shell detonation.  Model CSDD-L gives a value of 1.69
    for [Mn/Fe] from the helium shell detonation compared to 0.29 in
    Model CSDD-S.
  \item The progenitor metallicity plays an essential role for the
    production of Mn.
\end{enumerate}
Following argument (i) from above, a very low value of [Mn/Fe] could be
expected for Model M2a$_\odot$ because of the low-mass helium shell and
the $\sim1\, \msun$ core. The data, in contrast, show an approximately
solar value. The reason is that it has been calculated at solar
metallicity, which leads to [Mn/Fe]$_\mathrm{core} = -0.07$ (comparable
to Model VM) for core material and [Mn/Fe]$_\mathrm{shell}= 0.81$ for
the shell. Since the production of nuclei more neutron-rich than
$^{56}$Ni is enhanced for lower $Y_e$, this shift in [Mn/Fe] is
reasonable. The yields of model M2a$_\mathrm{Ne}$ fall into line with
this analysis. The [Mn/Fe] value in the core material, however, slightly
decreases indicating a low dependency on the initial distribution of
nuclei. Furthermore, Model M2a at zero metallicity from
\citet{gronow2020a} gives [Mn/Fe]$ = -1.00$, which supports the
explanations given above.

It is generally found that CO detonations at zero metallicity (also the
core detonations of CSDD-L and CSDD-S) produce less Mn than models at
higher metallicity (VM, M2a$_\odot$, M2a$_\mathrm{Ne}$).  In addition,
the pure detonation models (PD081, PD115) not only underproduce Mn with
respect to Fe (like VM) but also eject a total amount of Mn roughly
three orders of magnitude below all other explosions. Although we
neither have solar counterparts of CSDD-L, CSDD-S, PD081, PD115 nor zero
metallicity versions of VM, the comparison between the individual CO
detonations still indicates a metallicity-dependent Mn production.

In addition, some Mn is made via $^{55}$Fe directly only in the \mch
models. In Fig.~\ref{fig:co55} one can observe that $^{55}$Co is not
produced at the very tip of the high density end of the tracer particle
distribution in Model R60. These are exactly the conditions where the
even more neutron rich element $^{55}$Fe is synthesized.  In contrast to
M2a$_\mathrm{Ne}$, we do not observe any changes in the yields of
R60$_\mathrm{Ne}$ compared to R60. This is most likely due to the fact
that almost all of $^{55}$Co and $^{55}$Fe is synthesized in normal
freeze-out from NSE. Thus, the products largely depend on the
neutronization due to electron captures during the explosion phase (see
Sec.~\ref{subsec:electroncap}) and only weakly on the initial
metallicity.

It should be noted that explosions such as M2a do not resemble normal
SNe~Ia in some aspects (see also \citealp{fink2010a,gronow2020a}). A
cure to this problem might be to further decrease the mass of the helium
shell \citep{townsley2019double}, but this also reduces the production
of Mn in its detonation.  However, explosions such as HeD-S are a
candidate for Ca-rich transients, a sub-luminous class of SNe residing
between normal SNe~Ia and classical novae in terms of absolute
magnitude.  \citet{frohmaier2018b} estimate rather high rates for
Ca-rich transients of about 33\%-94\% of the rate of normal SNe~Ia. If
this proves to be correct, such explosions may substantially contribute
to the production of Mn in the Universe.

To estimate the effect of Ca-rich SNe on the evolution of [Mn/Fe] in the
Milky Way we carried out a chemical evolution simulation using the
One-Zone Model for the Evolution of Galaxies code \textsc{OMEGA}
\citep{cote2017a}. We calculate an open-box model and employ the
``star-formation model'' described by \citet{cote2017a} to control the
in- and outflows of gas.  These are linked to the star formation rate
$M_\star$ via a mass loading factor $\eta$. The total mass of gas inside
the galaxy is determined by the star formation efficiency $f_\star$:
$M_\star= f_\star M_\mathrm{gas}$. The star formation rate is taken from
\citet{chiappini2001a}, yields for massive stars are from
\citet{limongi2018a} (we use averaged values of their different rotating
and non-rotating models) and yields of AGB stars are extracted from
\citet{karakas2010a}. We find reasonable agreement with observational
data for Mn using different SN~Ia scenarios (see below) and fixing the
star formation efficiency to $f_\star=0.006$, the mass loading factor to
$\eta=0.7$, and the proportional constant connecting the star formation
time scale and the dynamical time scale to $f_\mathrm{dyn}=0.004$.
Moreover, the mass transition between AGB yields and massive star yields
is chosen to be $10.5\,M_\odot$. Finally, the rate of SNe~Ia is chosen
to $1.3 \times 10^{-3}\,M_\odot ^{-1}$, and the total number of SNe~Ia
is distributed according to the chosen contribution of each channel (see
cases below).

For the Chandrasekhar mass delayed detonations (N100ddt), helium-shell
double-detonations (CSDD-L, HeD-S), and violent WD  mergers (VM), we use
delay time distributions calculated with the {\sc StarTrack} binary
evolution code \citep[e.g.][]{belczynski2008a,ruiter2009a}.  For this
work, we assume Chandrasekhar mass exploding CO WDs that have a
hydrogen-rich donor (in most cases a subgiant or giant star) produce
delayed-detonations.  For violent WD mergers, we include any merger
between two CO WDs that has at least one component WD mass $\ge 0.9$
\msun. For double-detonations with helium shells, we employ the WD
mass-dependent helium shell prescription of \citet{ruiter2014a}.

\begin{figure}[htbp]
    \centering
    \includegraphics[width=\columnwidth]{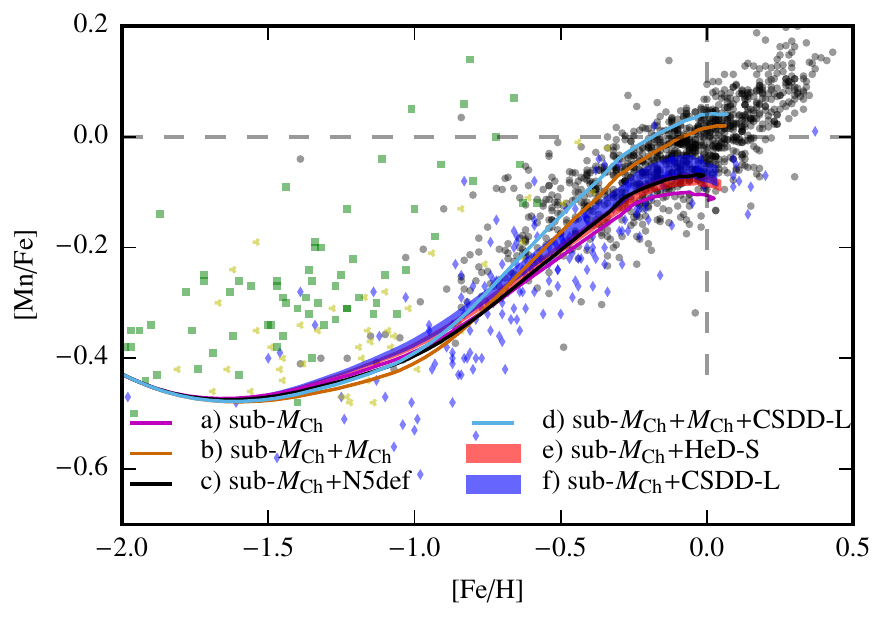}
    \caption{[Mn/Fe] evolution for different SNe~Ia scenarios. The
      sub-\mch scenario is represented by the Model VM, the \mch
      scenario by the N100ddt and N5def is used as a proxy for SNe
      Iax. The shaded areas in the runs sub-\mch+CSDD-L and
      sub-\mch+HeD-S correspond to a range in the rate of Ca-rich
      transients between 33\% and 94\%. Black dots show observational
      data of \citet{adibekyan2012b}, blue diamond shapes depict data
      of \citet{reddy2006a}, yellow triangles represent data of
      \citet{gratton2003a}, and red squares show data of
      \citet{ishigaki2012a, ishigaki2013a}. The data has been
      extracted from the \textsc{STELLAB} (STELLar ABundances) library
      \citep{2016ascl.soft10015R}. The non-LTE data by
      \citet{eitner2020a} are not included in this simple approach.}
    \label{fig:gce_mn}
\end{figure}

We implemented the following combinations of SN~Ia
channels. Combinations (a), (b), and (c) have also been investigated
by \citet{seitenzahl2013b}:
\begin{enumerate}[(a)]
  \item sub-\mch: 100\% Model VM
  \item sub-\mch + \mch: 50\% Model VM + 50\% Model N100ddt
  \item sub-\mch + Iax: 80\% Model VM + 20\% N5def
  \item sub-\mch + \mch + CSDD-L: same as case (b) with 50\% of
    \mbox{CSDD-L} added on top. This increases the number of SNe~Ia
    per stellar mass formed to $1.95\,M_\odot^{-1}$.
  \item sub-\mch + HeD-S: the SN~Ia rate consists 100\% of Model VM
    and 33\%-94\% of Model HeD-S are added on top
  \item sub-\mch + CSDD-L: same as case (e) with Model CSDD-L instead
    of HeD-S
\end{enumerate}
Combinations (a), (b), and (c) confirm the results described
above. The solar value of [Mn/Fe] can only be reached when including a
significant fraction of \mch explosions. Also a rather high fraction
of 20\% of failed deflagrations (N5def) represented by case (c) is not
sufficient to match the increasing trend.  However, we emphasize that
the upward trend in [Mn/Fe] is not only due to the contribution of
\mch-mass SNe~Ia. CCSNe also exhibit an increasing trend in [Mn/Fe]
with increasing metallicity. This is illustrated scenario (a) in which
SNe~Ia have virtually no contribution to Mn. However, despite the
CCSN-caused increase in Mn, a solar ratio is not reached.
Moreover, we find that helium detonations (HeD-S, CSDD-L) are as
effective in increasing the final value of [Mn/Fe] as N5def with our
choice of rates (see variations (e) and (f)). Furthermore, case (d)
demonstrates that CSDD-L also raises [Mn/Fe] in the presence of
N100ddt. This indicates that Ca-rich transients are able to reduce the
need for \mch explosions to reproduce the evolution of [Mn/Fe]. We
find that a reduction of \mch explosions to 30\% of the SN~Ia rate in
scenario (d) yields similar [Mn/Fe] values as in case (b). Replacing
CSDD-L with HeD-S allows a reduction to 40\%. Thus, the occurrence
rate of \mch SNe~Ia needed to explain [Mn/Fe] vs.\ [Fe/H] can be
reduced but not eliminated by the consideration of helium detonation
models.

We emphasize that this simple approach cannot replace future more
elaborate GCE studies. It is only intended to give an impression of the
contribution of helium shell detonation models compared to other SNe~Ia
scenarios given their low ejected mass per event.  The delay-time
distribution for explosions such as CSDD-L and HeD-S is not very well
constrained, for instance. Fortunately, this does not challenge our
conclusion since different DTDs only alter the shape of the [Mn/Fe]
evolution and leave the final value at [Fe/H] = 0 largely unaffected.
While there is still much uncertainty associated with the evolution and
explosive outcome of helium shell double-detonation binaries, we note
that in our binary evolution models progenitors of double-detonation
SN~Ia explosions are more similar in physical configuration to the
models of \citet{gronow2020a} (in terms of core and shell mass), rather
than the earlier models computed by \citet{sim2012a}. However, it turns
out that regardless of whether low to moderate shell mass systems
(Gronow) or high shell mass systems (Sim) are actually contributing to
the SN~Ia popualtion in Nature, it will not have any noticeable effect
on the delay-time distribution of these explosions, since the timescale
on which these WDs accumulate helium is by comparison insignificant (on
the order of ${\sim}10$ Myr).

In addition, a large number of helium detonations would lead to tensions
in [Ti/Fe] and [V/Fe] since these elements are produced in super-solar
amounts in them (see Fig.~\ref{fig:ratio_fe}). This is yet another
argument excluding helium shell detonations or double detonations as a
replacement for \mch SNe~Ia models producing a solar Mn/Fe ratio. As Ti
and V are, however, under-produced over the whole metallicity range in
current GCE studies (see \citealp{prantzos2018a} and references therein)
their site of production is not clarified completely, yet. The yields
derived from helium detonation models do not solve the problem. They
imply an increased production at $\mathrm{[Fe/H]} \approx 0$ and fail to
provide a good fit to the [Ti/Fe] and [V/Fe] evolution at low
metallicities. We find that a decrease of the CSDD-L rate down to
approximately 10\% of the SN Ia rate resolves this tension and yields
the solar value for [Ti/Fe] and [V/Fe] in scenario d). This, however,
only allows for a reduction of the \mch events to 45\%  compared to 30\%
mentioned above.

\subsection{Zinc}
\label{subsec:zinc}

\begin{figure*}[htbp]
    \centering
    \includegraphics[width=\textwidth]{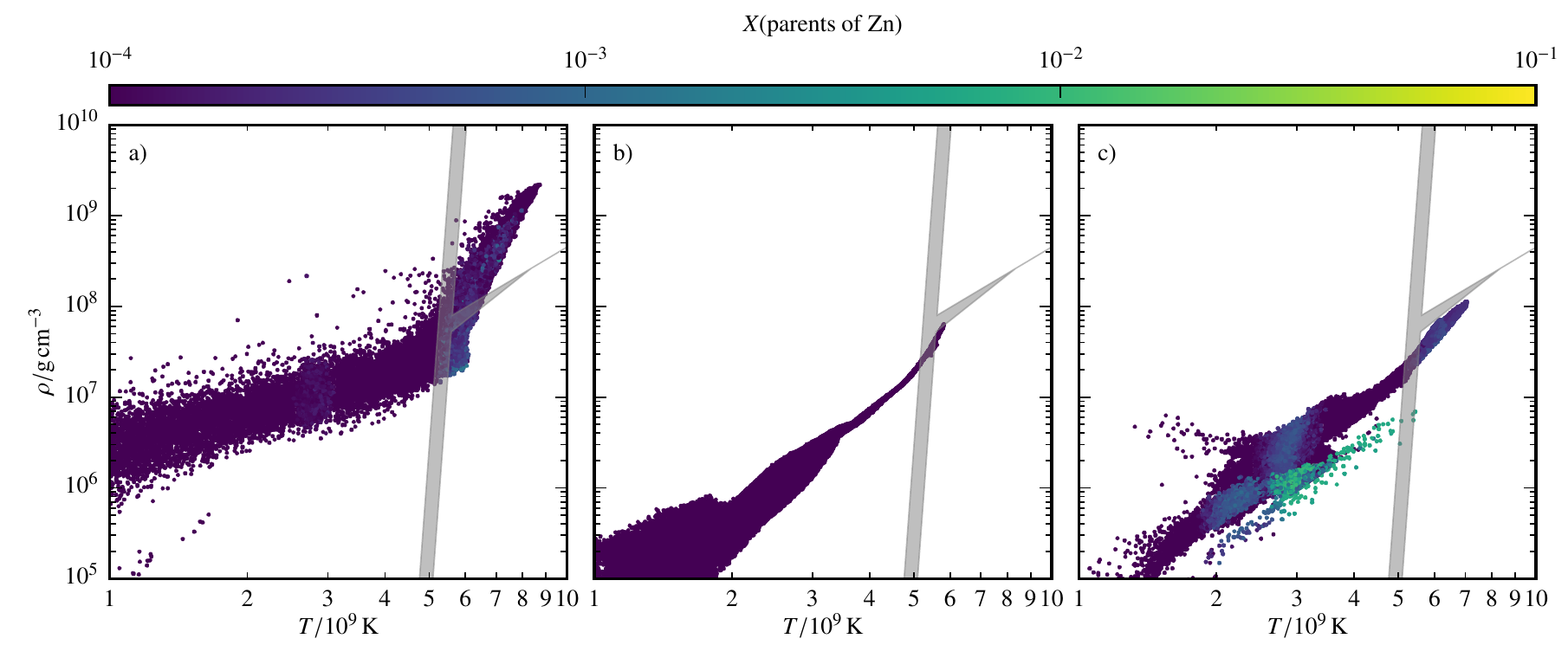}
    \caption{Same as Fig.~\ref{fig:ni56} with a color coded mass fraction of the parent nuclei of Zn
    listed in Table~\ref{tab:zinc_production}. The panels show: a) R60, b) VM, c)
  M2a$_\odot$}
    \label{fig:ge64}
  \end{figure*}

\begin{table*}
  \centering
  \caption{Total amount of stable Zn in $M_\odot$ and the
    radioactive and stable isotopes from which Zn 
    originates (in percent) at 100$\,$s. Only contributions larger than 1$\,$\%
  are listed.}
 \begin{tabularx}{\textwidth}{c @{\extracolsep{\fill}} ccccccccccc}
    \toprule
    model  &   Zn  & $^{64}$Zn  & $^{66}$Zn & $^{68}$Zn &  $^{64}$Ga & $^{64}$Ge &
    $^{66}$Ge & $^{67}$Ge & $^{68}$Ge & $^{67}$Cu  & $^{66}$Ni \\
    \midrule
    PD081  & 3.49e-16 & 97.9 &  -  &  -  & 1.5 &  -   &  -  &  -  & -  & - & -  \\ 
    PD115  & 5.42e-05 & 99.1 &  -  &  -  &  -  &  -   &  -  &  -  & -  & - & - \\ 
    VM     & 2.21e-06 & 1.6  &  -  &  -  & 9.4 & 28.3 & 60.6 & -  & -  & - & - \\ 
    M2a$_\odot$& 1.71e-04 & 4.2  & 10.3 & - &  7.8 & 39.1 & 34.0 & 1.2 & 2.1  & - & - \\
    M2a$_{\mathrm{Ne}}$ & 1.35e-04 & - & - &  - & 11.1 & 58.4 & 29.1 & -  &  - & - & - \\
    M2a    & 4.44e-04 &  -   &  -  &  -  & 13.0 & 79.9 &  5.4 &  -  &  -  & - & - \\
    CSDD-L & 6.05e-06 & 1.6  &  -  &  -  & 12.7 & 21.2 & 36.7 & 12.7 & 14.4 & - & - \\
    CSDD-S & 7.49e-04 &  -   &  -  &  -  &  7.0 & 85.4 & 6.9 & -  &  -  & - & - \\
    HeD-S  & 7.41e-04 &  -   &  -  &  -  & 6.8 & 85.5 & 7.0 & -  & -  & - & - \\
    N5def  & 6.58e-07 & 8.5  &  -  &  -  & 6.0 & 28.5 & 56.2 & -  &  -  & - & - \\ 
    N100ddt& 3.12e-06 & 5.6  &  -  &  -  & 5.9 & 30.9 & 57.2 & -  &  -  & - & - \\ 
    R60    & 1.80e-07 & 4.7 & 2.9 &  - & 4.1 & 33.6 & 53.0 & -  &  - & - & - \\ 
    R60$_{\mathrm{Ne}}$  & 1.92e-06 & - & - &  - & 3.9 & 31.6 & 49.6 & -  &  - & - & 14.1 \\ 
    W7     & 4.93e-07  & -    &  -  &  -  &  -  &  -   &  -  &  -  &  -  & - & - \\ 
    \bottomrule
  \end{tabularx}
  \label{tab:zinc_production}
\end{table*}

\begin{table*}
  \centering
  \caption{Zn to Fe ratio compared to solar for the total explosion as
  well as shell and core material only and stable zinc isotopes decayed
at $2\times 10^9\,$yr.} \begin{tabularx}{\textwidth}{c
    @{\extracolsep{\fill}} ccccccc}
    \toprule
    model  & [Zn/Fe] & [Zn/Fe]$_\mathrm{shell}$ & [Zn/Fe]$_\mathrm{core}$ & $^{64}$Zn  &  $^{66}$Zn  &  $^{67}$Zn  &  $^{68}$Zn \\ 
    \midrule
    PD081  &  -10.54 &  -  &  -  &  3.48e-16  &   -         &    -        &       -    \\
    PD115  &  -1.31 &  -  &  -  &  5.38e-05   &  3.57e-07   &    -        &       -    \\
    VM     &  -2.60 &  -  &  -  &  8.67e-07   &  1.34e-06   &    -        &       -    \\
    M2a$_\odot$& -0.66 & 0.82 & -0.92 & 8.72e-05  &  7.59-05   & 2.48e-06  & 5.11e-06  \\
    M2a$_{\mathrm{Ne}}$ & -0.77 & 0.77 & -1.09  & 9.46e-05  & 3.92e-05   & 5.60e-07 &  3.45e-07  \\
    M2a    &  -0.25 & 1.02 & -0.48 & 4.16e-04  &  2.46e-05   &   1.52e-06  &  1.64e-06  \\
    CSDD-L &  -0.71 & 0.07 & -1.77 & 2.14e-06  &  2.25e-06   &   7.86e-07  &  8.73e-07  \\
    CSDD-S &   0.40 & 0.87 & -1.44 & 6.94e-04  &  5.19e-05   &    -        &       -    \\
    HeD-S  &   0.87 & 0.87  &  -  & 6.86e-04  &  5.18e-05   &    -        &       -    \\
    N5def  &  -2.64 &  -   &  -  &  2.83e-07   &  3.74e-07   &    -        &       -    \\
    N100ddt&  -2.51 &  -   &  -  &  1.32e-06   &  1.79e-06   &    -        &       -    \\
    R60  &  -1.20 &  -   &  -  & 7.67e-07   &  1.02e-06   & 5.76e-09  &  8.67e-09  \\
    R60$_{\mathrm{Ne}}$  &  -1.17 &  -   &  -  & 6.96e-07   & 1.22e-06   & 1.48e-09  &  3.80e-10  \\
    W7     &  -1.66 &  -   &  -  &    -        &     -       &     -       &     -    \\
    \bottomrule
  \end{tabularx}
  \label{tab:stable_zinc}
\end{table*}

The element zinc ranges right beyond the iron peak and is of high
interest for GCE since its origin has not been clarified yet. It has
four stable isotopes, namely $^{64}$Zn, $^{66}$Zn, $^{67}$Zn and
$^{68}$Zn, of which $^{64}$Zn is the most abundant in the solar
neighborhood. The production mechanisms are therefore more diverse than
in the case of Mn. Zn abundances in the Galaxy have been measured
already by \citet{sneden1991b} and later on by \citet{mishenina2002c},
\citet{cayrel2004a}, and \citet{nissen2007b}. All agree on its
evolution: Zn exhibits high values of $\mathrm{[Zn/Fe]}\approx 0.6$ at
$\mathrm{[Fe/H]}\approx -4$, which drop to solar at around
$\mathrm{[Fe/H]}\approx -2$. From there on, they oscillate around
$\mathrm{[Zn/Fe]}\approx 0$. Some of the Zn abundance is synthesized
during He or C burning via the s-process in massive stars but the major
contribution comes from explosive nucleosynthesis in supernovae.
Standard CCSNe models \citep[see, e.g., the yields of][]{woosley1995a}
fall short in the production of Zn. Only the introduction of hypernovae
can account for the solar value of [Zn/Fe] \citep{kobayashi2006a}, but
the high values at very low metallicities are not reached within this
model either.

The most commonly used SN~Ia model for GCE calculations is the W7 model
\citep{iwamoto1999a} -- a 1D pure deflagration in a $M_\mathrm{Ch}$ CO
WD. This particular simulation yields only negligible amounts of Zn (see
Table~\ref{tab:zinc_production}). Therefore, it has been a goal to
explain the evolution of Zn with other production sites, although 1D
double detonation and pure helium detonation models calculated by
\citet{woosley2011b} show substantial amounts of Zn.  The recent work by
\citet{hirai2018a} tries to explain the evolution of Zn with
metallicity-dependent yields of CCSNe as well as HNe from
\citet{nomoto2013b}.  They find that the inclusion of electron capture
supernovae \citep[ECSNe, yields from][]{wanajo2018a} is necessary to
match the high [Zn/Fe] values at low metallicity. \citet{jones2019a}
presented nucleosynthesis yields of thermonuclear ECSNe (tECSNe), \ie
the explosion of ONe WDs at densities around $10^{10}\,\si{g.cm^{-3}}$.
These models overproduce neutron-rich isotopes such as $^{48}$Ca,
$^{50}$Ti, $^{54}$Cr as well as $^{66}$Zn and other elements beyond the
iron peak. In a follow up study \citet{jones2019b} showed that these
models complement nicely the contribution of ECSNe to the chemical
evolution of the Milky Way. \citet{prantzos2018a} cannot reproduce the
evolution of Zn using the nucleosynthesis yields of rotating massive
stars by \citet{limongi2018a}.

In spite of this, a high contribution of SNe~Ia to Zn has already been
proposed by \citet{matteucci1993b} and later on by
\citet{francois2004a}. \citet{mishenina2002c} claim that SNe~Ia are
responsible for as much as 67\% of the Zn production.
\citet{tsujimoto2018a} investigate the evolution of Zn in the Galaxy
using Mg instead of Fe as the reference element. Mg is an
$\alpha$-element assumed to be solely produced in CCSNe and thus it is
more sensitive in detecting the contribution of sources other than CCSNe
to a specific element. They discover a decreasing trend for [Zn/Mg] for
$\mathrm{[Zn/Mg]}\lesssim -1$ and a rise for higher metallicities. This
increasing behavior coincides with the well known kick-in of SNe~Ia at
$\mathrm{[Fe/H]}=-1$. Consequently, they conclude that SNe~Ia must be
responsible for this behavior and suggest a scenario including a He
detonation with strong $\alpha$-rich freeze-out. The decreasing trend at
low metallicities is explained in their GCE model by the incorporation
of magnetorotational SNe (MR SNe), whose rate decreases with increasing
metallicity.

No matter which combination of supernova scenarios accounts for the
observed abundances at $\mathrm{[Fe/H]}\lesssim -1$, it is very likely
that SNe~Ia also contribute to Zn in a non-negligible way, \ie a
significant underproduction would require an even larger contribution
from CCSNe to keep [Zn/Fe] near the solar value.  The question is which
scenario for SNe~Ia synthesizes [Zn/Fe] at around the solar ratio or
higher and therefore contributes to the enrichment of galaxies with Zn?

Table~\ref{tab:zinc_production} lists the total production of Zn in
solar masses as well as the fraction of isotopes via which it is
produced.  In addition, Table~\ref{tab:stable_zinc} shows [Zn/Fe] for
the whole explosion as well as for the helium shell and core detonation
separately in the case of a double detonation. It reveals that all
models, except for those including a helium detonation (M2a$_\odot$,
M2a, CSDD-S, CSDD-L, and HeD-S), severely underproduce [Zn/Fe] compared
to its solar value. While M2a$_\odot$, M2a$_\mathrm{Ne}$, M2a, and
CSDD-L exhibit only a moderate underproduction, the remaining models
even synthesize super-solar ratios [Zn/Fe]. The [Zn/Fe] value of the
helium detonation is quite similar in M2a$_\odot$, CSDD-S and HeD-S but
the result of the whole explosion is dominated by the core detonation
for model M2a$_\odot$. In contrast to the case of Mn, the values of
[Zn/Fe] in M2a are higher than its solar metallicity counterpart. The
most important production channel is via the symmetric nucleus $^{64}$Ge
as is also the case for CSDD-S and HeD-S.  In Model CSDD-L, the [Zn/Fe]
yield of the helium shell detonation is significantly lower than in the
other models. The reason for this is the lower density of the envelope
(see argument (ii) in Sect.~\ref{subsec:manganese}) because most Zn is
produced above an initial density of $5.0\times 10^5\,\gcc$. The lower
density is also responsible for the difference in the contribution of
$^{64}$Ge compared to the other double detonations as it is produced at
higher densities than, e.g., $^{66}$Ge.

Furthermore, it can be seen that the main production channels are
$^{64}$Ga$\to^{64}$Zn, $^{64}$Ge$\to^{64}$Ga$\to^{64}$Zn and
$^{66}$Ge$\to^{66}$Ga$\to^{66}$Zn for the majority of models except for
PD081 and PD115. These isotopes are produced either in $\alpha$-rich
freeze-out from NSE or in the helium detonation (see
Fig.~\ref{fig:ge64}).

Fig.~\ref{fig:ge64} also shows that Zn is primarily produced in the
helium shell at relatively high densities. This region is not reached by
CSDD-L as discussed above. A direct production of Zn in the form of a
high contribution from $^{64}$Zn can only be observed in the pure
detonations PD081 and PD115 and -- to a much lesser extent -- in the
$M_\mathrm{Ch}$-models. However, the total amount of Zn falls short of
that of Models CSDD-S, HeD-S, and M2a$_\odot$ by about two orders of
magnitude.  R60 and M2a$_\odot$ are the only models to directly produce
a non-negligible amount of $^{66}$Zn. Moreover, the direct production of
Zn isotopes ($^{64,66}$Zn) is clearly a metallicity effect. The
corresponding simulations at lower metallicity shift the production of
Zn to the symmetric isotope. This behavior is not observed for the other
models at solar metallicity (N5def, N100ddt, VM). In contrast to our
most recent simulations, \ie M2a$_\odot$ and R60, where the solar value
for each isotope according to \citet{asplund2009a} is used as input for
the postprocessing, the metallicity is set by adding only $^{22}$Ne to
adjust the electron fraction. Therefore, the lack of seed nuclei in the
investigated region might affect the detailed nucleosynthesis results.
This is confirmed by models R60$_\mathrm{Ne}$ and M2a$_\mathrm{Ne}$.
They do not produce any Zn isotopes and also show differences in the
production of various other species.

In summary, any SN~Ia scenario noticeably contributing to the enrichment
of the Galaxy with Zn is required to include a prominent helium
detonation. The production of Zn can be explained by the same three
arguments used for the case of Mn. However, here CSDD-L shows the lowest
ratio to iron since the initial shell density is too low to synthesize
Zn in sufficient amounts. We abstain from adding an investigation of the
galactic evolution of [Zn/Fe] since we have already shown for the case
of Mn that low-mass double detonations and pure helium shell detonations
do contribute significantly to [Mn/Fe] despite their low ejecta masses.
This result holds analogously for [Zn/Fe].

Instead, we briefly estimate the influence of $^{68}$Ge on the light
curve. All other isotopes listed in Table~\ref{tab:zinc_production} are
short-lived with a maximum half-life\footnote[1]{All nuclear decay data
is taken from https://www.nndc.bnl.gov/nudat2/} of $61.8\,\mathrm{h}$
for $^{67}$Cu. Although compared to $^{56}$Ni long-lived isotopes are
not produced in large amounts, they can modify the shape of the light
curve at late times \citep{seitenzahl2009d}. Furthermore, $\gamma$-rays
or X-rays emitted in their decays might be detectable.

$^{68}$Ge decays to $^{68}$Ga via electron capture with a half-life of
$T_{1/2} =270.95\,$d. The two most relevant X-rays emitted have energies
of $E_1=9.225$ and $E_2=9.252\,$keV with an emission probability of
$I_1=0.131$ and $I_2=0.258$, respectively. As an example, we consider
Model M2a$_\odot$, which ejects $3.554 \times 10^{-6}\,\msun$  of
$^{68}$Ge.  Assuming transparent ejecta after $t=100\,$d and a very
close SN explosion at a distance of $d=1\,$Mpc we arrive at a flux on
Earth of 
\begin{align}
  F_i &= \frac{1}{4\pi d^2} \cdot I \lambda N_0 e^{-\lambda t} \cdot E_i = \nonumber \\
  &=
  \begin{cases}
    2.325\times 10^{-17} \si{erg.s^{-1}.cm^{-2}} & i=1 \\
    4.592\times 10^{-17} \si{erg.s^{-1}.cm^{-2}} & i=2 \\
  \end{cases},
  \label{eq:fluxge68}
\end{align}
where $\lambda$ is the decay constant and $N_0$ the number of nuclei at
$t=0$, which can be derived dividing the ejected mass by the
corresponding atomic mass\footnote[2]{The atomic mass is taken from
https://www-nds.iaea.org/}.  The X-ray telescope \textsc{NuSTAR}
\citep{harrison2013a} has a sensitivity of $2\times
10^{-15}\si{erg.s^{-1}.cm^{-2}}$ in the considered energy range.
Therefore, the added flux of the two X-ray emissions is about a factor
of $35$ below the detection limit, and thus, a higher production of
$^{68}$Ge would be necessary for a potential detection even under the
rather favorable conditions assumed here. We emphasize that this
statement is by no means conclusive since the sensitivity of
\textsc{NuSTAR} also depends on the line shape and on the observation
time. 

\begin{figure}[htbp]
  \centering
  \includegraphics{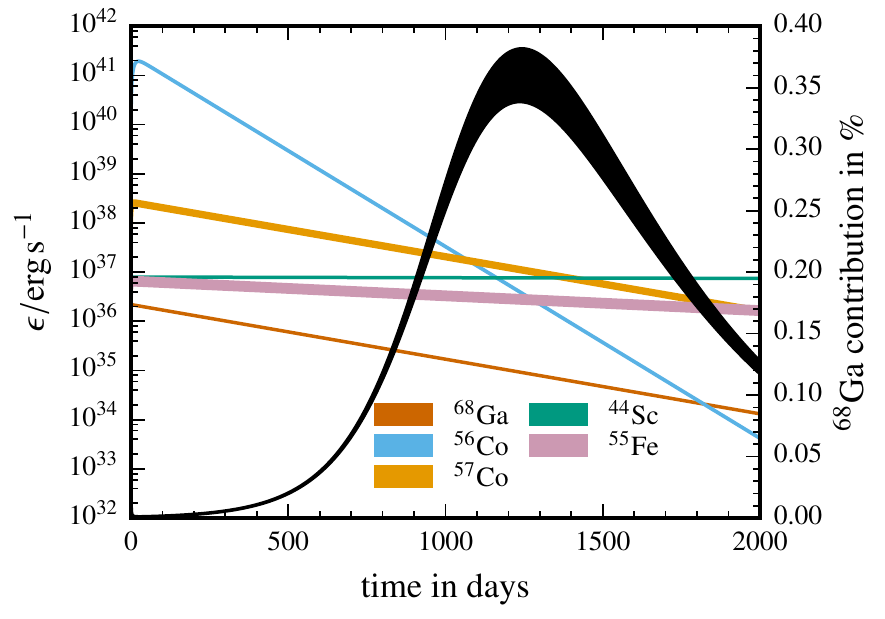}
  \caption{Energy generation rate $\epsilon$ of the emission of positrons,
    conversion electrons, and Auger electrons in model M2a$_\odot$.
    Shaded areas show the range between no X-ray trapping and full X-ray
    trapping. The black curve depicts the contribution of the
    $^{68}$Ga decay relative to the total energy generation. }

  \label{fig:covsga}
\end{figure}

Subsequently, $^{68}$Ga decays to $^{68}$Zn via positron emission very
quickly ($T_{1/2} = 67.71\,\mathrm{min}$). Assuming an instantaneous
energy deposition, we calculated the contributions of $^{68}$Ga,
$^{44}$Sc, $^{56}$Co, $^{57}$Co, and $^{55}$Fe positrons, conversion
electrons, and Auger electrons to the SN light curve for Models HeD-S
(same amount of $^{68}$Ge as CSDD-S), CSDD-L and M2a$_\odot$.  In the
case of $^{68}$Ga and $^{44}$Sc, the electrons emitted by their
long-lived parent are included, too. Fig.~\ref{fig:covsga} shows that
with less than $0.4\%$ of the energy generation per second $\epsilon$,
the $^{68}$Ga decay plays only a minor role in model M2a$_\odot$.  Its
contribution is even lower in the HeD-S and CSDD-L models. 

\subsection{Copper}
\label{subsec:copper}

\begin{figure*}[htbp]
    \centering
    \includegraphics[width=\textwidth]{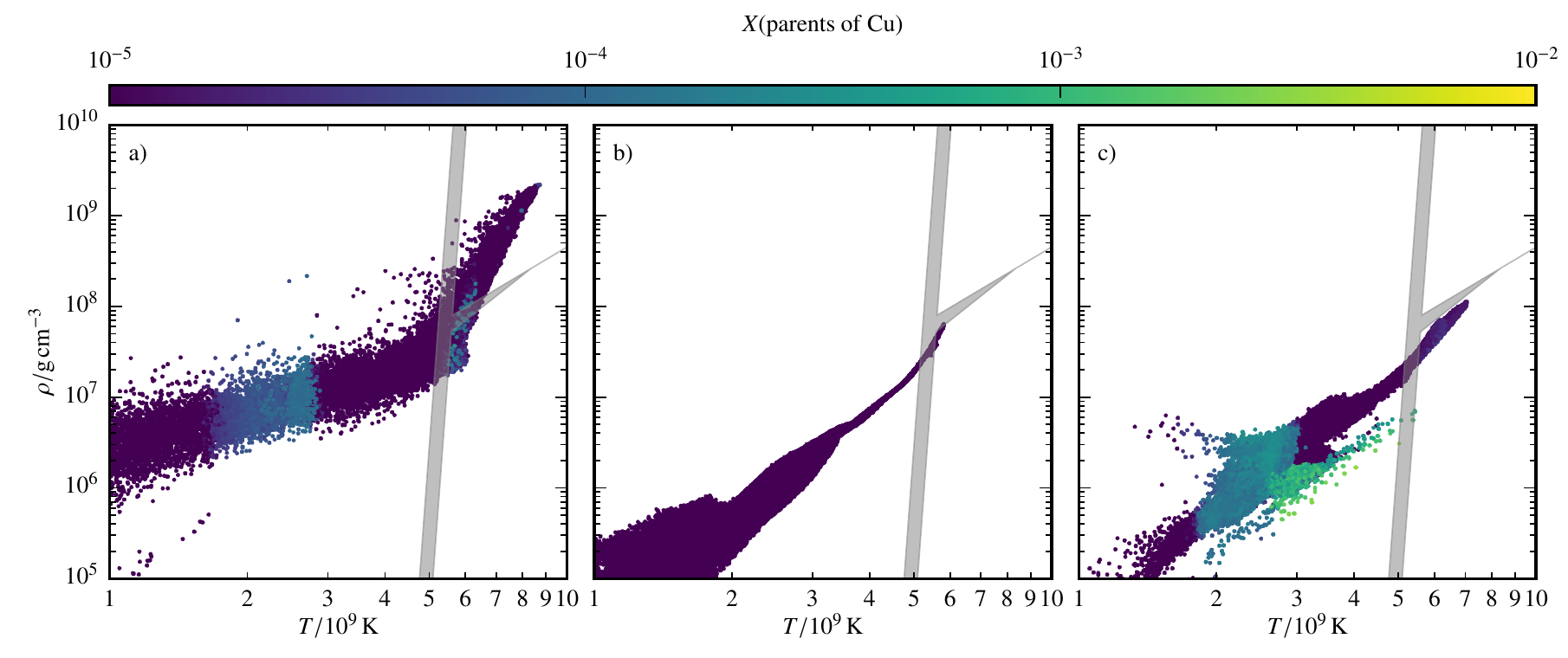}
    \caption{Same as Fig.~\ref{fig:ni56} with a color coded mass
    fraction of $^{64}$Ge. The panels show: a) R60, b) VM, c)
    M2a$_\odot$}
    \label{fig:ga63}
\end{figure*}

\begin{table*}
  \centering
  \caption{Total amount of stable Cu in $M_\odot$ and the
    radioactive and stable isotopes from which Cu 
    originates (in percent) at 100$\,$s. Only contributions larger than 1$\,$\%
    are listed.}
 \begin{tabularx}{\textwidth}{c @{\extracolsep{\fill}} ccccccccccc}
    \toprule
    model  & Cu & $^{63}$Cu & $^{65}$Cu & $^{63}$Zn & $^{65}$Zn & $^{63}$Ga & $^{65}$Ga &
    $^{65}$Ge & $^{63}$Ni & $^{65}$Ni & $^{63}$Co \\
    \midrule  
    PD081  &  1.66e-15 & 33.4 & -    & 66.0 & - & -   &  -    &  -  &   -    & -  & -  \\ 
    PD115  &  2.48e-05 & 34.2 & -    & 64.9 & - & -   &  -    &  -  &   -    & -  & -  \\ 
    VM     &  3.37e-07 & -    & -    & 74.2 & - & 8.8 & 13.9  & 2.0 &   -    & -  & -  \\ 
    M2a$_\odot$&  2.43e-05 & 10.6  & 10.8 & 15.5 & 3.6 & 10.5 & 33.4 & 12.6 & 2.2  & -  & - \\
    M2a$_{\mathrm{Ne}}$  &  1.60e-05 & - & -  & 20.2 & - & 16.9 & 45.1 &  17.3 & - & - & - \\ 
    M2a    &  3.03e-05 & -    & -    & 18.2 & 3.0 & 23.5& 41.9 & 14.1 & -    & -  & -   \\
    CSDD-L &  3.44e-06 & -    & -    & 27.0 & - & 16.0& 26.2  & 30.4&   -    & -  & -  \\
    CSDD-S &  1.29e-04 & -    & -    & 18.8 & - & 57.7& 9.9   & 13.6&   -    & -  & -  \\
    HeD-S  &  1.31e-04 & -    & -    & 17.9 &   -  & 59.1 & 9.4 & 13.6 & -  & -  & -  \\
    N5def  &  1.15e-07 & 38.9 & 1.4  & 30.8 & - & 8.3  & 9.5  &  4.5 &  6.7  & -  & -  \\ 
    N100ddt&  4.58e-07 & 29.6 &  -   & 38.9 & - & 10.1 & 11.5 &  5.0 &  3.8  & -  & -  \\ 
    R60    &  2.54e-07 & 22.3 & 8.3  & 9.3 & 3.1 & 3.5 & 12.5 &  16.6 & 13.3 & 2.3 & 8.9 \\ 
    R60$_{\mathrm{Ne}}$   &  1.13e-07 & 2.3  & -  & 20.7 & - & 7.8 & 28.1 &  37.2 & - & - & 2.4 \\ 
    W7     &  3.21e-07 & - & - & -& -& - & -\\ 
    \bottomrule
  \end{tabularx}
  \label{tab:copper_production}
\end{table*}

\begin{table*}
  \centering
  \caption{Cu to Fe ratio compared to solar for the total explosion as
  well as shell and core material only and stable copper isotopes
  decayed at $2\times 10^9\,$yr}
  \begin{tabularx}{\textwidth}{c @{\extracolsep{\fill}} ccccc}
    \toprule
    model  &  [Cu/Fe] & [Cu/Fe]$_\mathrm{shell}$ & [Cu/Fe]$_\mathrm{core}$ & $^{63}$Cu  &  $^{65}$Cu   \\ 
    \midrule
    PD081  &  -9.48 &  -  &  -  &  1.65e-15   &  9.78e-18   \\
    PD115  &  -1.25 &  -  &  -  &  2.55e-05   &  2.33e-07    \\
    VM     &  -3.03 &  -  &  -  &  2.82e-08   &  5.44e-08    \\
    M2a$_\odot$&  -1.13 & 0.43 & -1.48 & 9.62e-06   & 1.47e-05    \\
    M2a$_{\mathrm{Ne}}$ & -1.31 & 0.39 & -1.93 & 5.95e-06   &  1.01e-05    \\
    M2a    &  -1.04 & 0.43 & -1.51 & 1.29e-05   &  1.75e-05    \\
    CSDD-L &  -0.57 & 0.21 & -1.69 & 1.49e-06   &  1.95e-06    \\
    CSDD-S &   0.02 & 0.48 & -1.31 & 9.86e-05   &  3.02e-05    \\
    HeD-S  &   0.50 & 0.50 &  -  & 1.01e-04   &  3.01e-05    \\
    N5def  &  -3.01 &  -  &  -  & 9.76e-07   &  1.77e-08    \\
    N100ddt&  -2.95 &  -  &  -  & 3.78e-07   &  8.04e-08    \\
    R60    &  -1.67 &  -  &  -  & 1.45e-07   &  1.09e-07    \\
    R60$_\mathrm{Ne}$  & -2.02 &  -  &  -  & 3.84e-08   & 7.50e-08    \\
    W7     &  -1.46 &  -  &  -  &  -  &  -   \\
    \bottomrule
  \end{tabularx}
  \label{tab:stable_copper}
\end{table*}

Copper directly follows nickel in the periodic table and its elemental
solar abundance consists of two stable isotopes, namely $^{63}$Cu and
$^{65}$Cu. Of these, about 69\% are attributed to $^{63}$Cu
\citep{asplund2009a}. The first extensive measurements and an analysis
of Cu abundances have been carried out by \citet{sneden1991b}. The
general trend of an increasing value of [Cu/Fe] with [Fe/H] in the
galaxy has already been established in that work. [Cu/Fe] rises from a
value of approximately $-1$ at $\mathrm{[Fe/H]}\approx -3$ to the solar
value at $\mathrm{[Fe/H]}\approx 0$. 

As for Zn, the origin of Cu is still very uncertain. The largest
contributors are believed to be the weak component of the s-process in
massive stars (a secondary process) and the direct fusion as a primary
element simultaneous to IGEs in CCSNe and SNe~Ia.
\citet{matteucci1993b} carry out detailed GCE calculations. Comparing to
the \citet{sneden1991b} data they conclude that a best fit is achieved
if the SNe~Ia yields for Cu are raised by an order of magnitude.
Moreover, they claim that SNe~Ia start contributing to the enrichment of
the Galaxy already from [Fe/H]$\approx -1.7$, \ie already in the halo
phase.  \citet{mishenina2002c} provide a large upgrade to the Cu
abundance data by measuring Cu and Zn in 90 metal poor stars in the
Galaxy. They arrive at the conclusion that the increase in [Cu/Fe] is
due to a significant contribution of SNe~Ia. As a guideline they also
provide a very rough estimate of the relative contributions of Cu from
different production sites to the solar abundance. They assign 7.5\% to
SNe II (primary process in massive stars), 25\% to secondary processes
in massive stars, 5\% to the s-process in AGB stars and 62.5\% to
SNe~Ia. The overall trend of the Cu to Fe ratio, \ie a sub-solar plateau
at low metallicities and an increase to the solar value, which is
reached at $\mathrm{[Fe/H]}\approx-0.8$, has also been confirmed by
\citet{reddy2003a}. In contrast, \citet{prantzos2018a} show that by
including the yields of \citet{limongi2018a} the evolution of [Cu/Fe]
can be modeled with CCSNe alone (they use W7 for SNe~Ia). Also
\citet{nissen2011a} confirm sub-solar [Cu/Fe] values for their
low-$\alpha$ population of dwarf stars in the solar neighborhood. These
stars are suspected to be primarily enriched by SNe~Ia and therefore
they conclude that these are not a main contributor to Cu.
\citet{simmerer2003a} investigate the Cu abundances in various globular
clusters and draw the opposite conclusion -- that SNe~Ia are likely to
be the main contributors to Cu -- because [Cu/Fe] in globular clusters
follows the trend seen in field stars. However, the only cluster
spanning a significant range in metallicity ($-1.8
<\mathrm{[Fe/H]}<0.8$) is $\omega$-Centauri and there the [Cu/Fe] curve
is rather flat. This suggests a different chemical evolution history. In
contrast to the previously mentioned results, \citet{romano2007b} find
that their GCE model can fit both, the Galaxy and $\omega$-Centauri, if
the s-process yields are enhanced and the SN~Ia yields are reduced.
\citet{cunha2002b} investigate in more detail the evolution of
$\omega$-Centauri. They attribute the strong enhancement of
$\alpha$-elements relative to Fe and the constant evolution of [Cu/Fe]
at a rather low value of -0.5 to an enrichment via CCSNe. However, a
final explanation for the lack of SNe Ia enrichment is not given. Recent
non-LTE Cu abundance measurements are presented by
\citet{yan2015a,yan2016a}. They show that non-LTE effects raise the
previously measured values by about $0.2\, \mathrm{dex}$ and that this
difference is also metallicity-dependent. The impact of the non-LTE
treatment is higher for lower [Fe/H] and therefore flattens the whole
curve. Furthermore, the plateau at lower metallicities is extended
compared to older works, and reaches up to [Fe/H]$\approx -1$\, and then
rises to solar. However, \citet{yan2015a} emphasize that due to the
uncertainties in the main production site of Cu, the peculiar evolution
of Cu in Ursa Major, $\omega$-Centauri, the Sagittarius dwarf galaxy,
and the halo sub-population \citep{nissen2011a}, further GCE models
should be postponed until the trend of [Cu/Fe] is clearly established by
using non-LTE measurements.

Similar to the discussion on Zn (see Sect.~\ref{subsec:zinc}), it is now
interesting to see which SN~Ia scenario is able to synthesize Cu in
significant amounts. Table~\ref{tab:copper_production} displays the Cu
yields together with the relative contribution of the most important
production channels. In Table~\ref{tab:stable_copper} the [Cu/Fe] values
and the remaining stable Cu isotopes can be found.

The overall result is the same as for the Zn case: Cu is largely
underproduced compared to solar in all simulations except for two models
including a He detonation; in this case HeD-S and CSDD-S. However,
[Cu/Fe] values are lower than [Zn/Fe] for the helium shell, which also
diminishes the total [Cu/Fe] yields. M2a$_\odot$, M2a$_\mathrm{Ne}$ and
M2a even approach the pure detonation PD115 due to the low shell mass to
core mass ratio (argument (i) in Sect.~\ref{subsec:manganese}). A
comparison between M2a$_\odot$ and M2a shows that the metallicity has
little impact on the total value of [Cu/Fe]. It only slightly changes
the fractions of the parent nuclei that decay to stable Cu. In
particular, we find that the direct production of $^{63}$Cu and
$^{65}$Cu is favored for higher metallicity and that the inclusion of
only $^{22}$Ne leads to a decrease in [Cu/Fe] (see R60$_\mathrm{Ne}$,
M2a$_\mathrm{Ne}$). Moreover, the difference in [Cu/Fe] from the helium
shell between CSDD-L and the other double detonations is less
significant than for [Zn/Fe]. $^{63}$Ga, for example, is mainly produced
at an initial density higher than the density at the base of the shell
in CSDD-L and therefore the relative contribution to stable Cu is
shifted to $^{65}$Ga and $^{65}$Ge compared to Models CSDD-S and HeD-S.

In contrast to Zn, a main production channel for Cu cannot be
identified. While most models, except R60, produce some Cu via
$^{63}$Zn, the helium detonations additionally show a major contribution
from $^{63}$Ga, $^{65}$Ga and $^{65}$Ge. The \mch simulations and PD115
also synthesize considerable amounts of $^{63}$Cu and M2a$_\odot$
exhibits contributions from almost every parent isotope listed in
Table~\ref{tab:copper_production}. This reflects that neither the core
nor the helium detonation dominates the total Cu yields.

To sum up, the creation of a large amount of Cu in relation to Fe
requires an even more prominent helium detonation than in the case of
Zn. We find in addition that a density at the base of the helium
envelope exceeding that in Model CSDD-L of
$5.92\times10^{5}\,\si{g.cm^{-3}}$ is essential. From a GCE point of
view the contribution of helium detonations to Cu is less significant
than in the case of Mn or Zn among the investigated models.  The only
model expected to contribute considerably to [Cu/Fe] is HeD-S. Also in
this case we postpone a detailed GCE modeling to future work.

Similar to the discussion of long-lived radioactive isotopes in
Sect.~\ref{subsec:zinc} we shortly take a look at $^{65}$Zn. The only
other long-lived isotope in Table~\ref{tab:copper_production} is
$^{63}$Ni. However, it is only produced in very small amounts in model
R60 and its half-life of $101.2\,\mathrm{y}$ implies a very low rate of
decays.  $^{65}$Zn, in contrast, has a half-life of $T_{1/2} =
243.93\,$d and emits an energetic $\gamma$-ray with an energy of
$E=1115.539\,\si{keV}$ and an emission probability of $I=0.5004$. Most
of $^{65}$Zn is synthesized via the fast decaying isotopes $^{65}$Ge and
$^{65}$Ga (see Table~\ref{tab:zinc_production}). Under the same
assumptions as in Sect.~\ref{subsec:zinc} we arrive at a flux on Earth
for Model CSDD-S (highest amount of $^{65}$Zn among our models with
$3.019\times 10^{-5}\,\msun$) of \begin{align} F = 1.031 \times 10^{-13}
  \si{erg.s^{-1}.cm^{-2}} .  \label{eq:zn65flux} \end{align} Although
this value is four orders magnitude higher than for the $^{68}$Ge X-rays
(see Sect.~\ref{subsec:zinc}), it is still at least an order of
magnitude below the expected sensitivity of the planned $\gamma$-ray
telescope e-ASTROGAM \citep{deangelis2018a}, and therefore a detection
is very unlikely. 

\section{Summary}
\label{sec:summary}

We investigate the nucleosynthesis yields of a variety of SN~Ia models
from the \textsc{HESMA} \citep{kromer2017a} archive as well as two new
explosion simulations, M2a$_\odot$ and R60. These models include double
detonation models from \citet{sim2012a} and \citet{gronow2020a}.
Furthermore, we examined the pure detonations in sub-\mch WDs of
\citet{sim2010a} and a set of \mch explosions from
\citet{seitenzahl2013a}, \citet{fink2014a}.

We aim to identify elements characteristic for a certain explosion
mechanism. In combination with GCE calculations and abundance
measurements in stellar atmospheres this can help to identify SN~Ia
progenitors. A well known example is the element manganese. Its
abundance relative to iron increases from [Fe/H] $\approx -1$ and it is
not produced by CCSNe in sufficiently high ratios relative to Fe.
Therefore, it is attributed to thermonuclear explosions of \mch WDs
\citep{seitenzahl2013b}. In this study we find that super-solar amounts
of Mn are additionally synthesized in helium detonations and that the
actual result depends on the interplay of three parameters: the
metallicity, the helium shell mass compared to the CO core, and the
density of the helium envelope. A variation of these values gives rise
to sub-solar values of [Mn/Fe] (as seen in Models M2a and CSDD-S) up to
highly super-solar results (CSDD-L). This also brings the double
detonation scenario into play as a potential source of Mn. Therefore,
the sole distinction between sub-\mch and \mch mass models when
investigating the source of Mn is insufficient. The details of the
underlying model employed for the sub-\mch channel have to be defined in
addition. Our GCE calculation corroborates this result and demonstrates
that double detonation models with massive helium shells can
significantly reduce but not completely remove the need for \mch
explosions to explain the solar [Mn/Fe] ratio. Their actual rate,
however, is limited due to their super-solar [Ti/Fe] and [V/Fe] values. 

Moreover, the elements zinc and copper have received little attention
when studying the contribution of SNe~Ia to galactic chemical evolution,
although \citet{matteucci1993b} already pointed out the potential
relevance of these events as a production site. We find that double
detonation models are able to produce Zn and Cu in super-solar ratios
with respect to Fe. Since a sophisticated GCE modeling and a
comprehensive analysis of the contribution of massive stars is beyond
the scope of this work we did not include models for the evolution of
[Zn/Fe] and [Cu/Fe]. However, the general results from the case of Mn
carry over to Zn and Cu.  The same three parameters as in the
case of Mn also affect the creation of Zn and Cu: Essentially, any value
of [Mn/Fe], [Zn/Fe] or [Cu/Fe] can be realized using different
combinations of helium shell mass, core mass, shell density and
metallicity. However, in this study we find that the models from
\citet{sim2012a} affect the evolution of Mn, Zn and Cu the most. These
models follow explosions in a system of a low-mass CO cores covered by a
massive helium shell. They were intended to resemble a sub-luminous
sub-class of SNe~Ia, namely calcium-rich transients, but do not account
for normal SNe~Ia. Their significance for GCE therefore depends on the
realization frequency of Ca-rich transients, which is currently
afflicted with large uncertainties. Ongoing and planned transient
searches hold promise to clarify this aspect.

We thus emphasize that SNe~Ia, or, in general thermonuclear explosions,
should be treated more carefully in GCE studies. It seems to be
necessary to include a variety of thermonuclear explosion models rather
than sticking to only one or two scenarios. SNe~Ia should be considered
as a source of Zn and Cu in GCE simulations if the double detonation
scenario is used to represent either normal SNe~Ia \citep[see,
however][for potential problems with the predicted spectral
observables]{kromer2010a} or the faint class of calcium-rich transients.
\mch explosions might not be the only relevant source of Mn.

\begin{acknowledgements} This work was supported by the Deutsche
  Forschungsgemeinschaft (DFG, German Research Foundation) -- Project-ID
  138713538 -- SFB 881 (``The Milky Way System'', subproject A10), by
  the ChETEC COST Action (CA16117), and by the National Science
  Foundation under Grant No. OISE-1927130 (IReNA).  FL , FKR and SG
  acknowledge support by the Klaus Tschira Foundation.

IRS was supported by the Australian Research Council through Grant
FT160100028.  AJR was supported by the Australian Research Council
through Grant FT170100243.  NumPy and SciPy \citep{oliphant2007a},
IPython \citep{perez2007a}, and Matplotlib \citep{hunter2007a} were used
for data processing and plotting.  The authors gratefully acknowledge
the Gauss Centre for Supercomputing e.V.  (www.gauss-centre.eu) for
funding this project by providing computing time on the GCS
Supercomputer JUWELS \citep{juwels2019} at J\"{u}lich Supercomputing
Centre (JSC).  Part of this research was undertaken with the assistance
of resources and services from the National Computational Infrastructure
(NCI), which is supported by the Australian Government, through the
National Computational Merit Allocation Scheme and the UNSW HPC Resource
Allocation Scheme.

BC acknowledges support from the ERC Consolidator Grant (Hungary)
funding scheme (project RADIOSTAR, G.A. n. 724560) and from the National
Science Foundation (USA) under grant No. PHY-1430152 (JINA Center for
the Evolution of the Elements).

\end{acknowledgements}

\bibliography{astrofritz}
\bibliographystyle{aa}

\end{document}